\documentclass[letterpaper,twocolumn,10pt]{article}
\usepackage{usenix,epsfig,endnotes}
\usepackage{xcolor}
\usepackage{url}
\usepackage{enumitem}
\usepackage{balance}
\usepackage{tabularx}

\usepackage{breakurl}
\usepackage{color}
\usepackage{booktabs}
\usepackage{adjustbox}
\usepackage{multirow}
\usepackage[normalem]{ulem}
\useunder{\uline}{\ul}{}
\usepackage{color}
\newcommand{\citep}[1]{\cite{#1}}

\begin{document}

\date{}

\title{\Large \bf Ctrl-Shift: How Privacy Sentiment Changed from 2019 to 2021}

\author{
{\rm Angelica Goetzen}\\
Max Planck Institute for Software Systems\\
agoetzen@mpi-sws.org
\and
{\rm Samuel Dooley}\\
University of Maryland\\
Max Planck Institute for Software Systems\\
sdooley1@cs.umd.edu
\and
{\rm Elissa  M. Redmiles}\\
Max Planck Institute for Software Systems\\
eredmiles@mpi-sws.org
}

\maketitle

\section{Introduction}
Privacy, as defined by Westin~\cite{westin2003social}, is an individual’s right to determine what personal information should be known or used by others. While the meaning and importance of privacy can vary by context~\cite{westin2003social,dienlin2014privacy}, national polls taken across 15 years show that a majority of Americans view privacy as an important right~\cite{best2006privacy}. Polls over the years have captured shifts in sentiments towards different areas of privacy \cite{westin2003social,katz1990report,best2006privacy}, with recent data showing Americans expressing concern over the way their information has been used by private companies and the government~\cite{auxier2019americans}.

Shifts in privacy sentiment have been observed in tangent with impactful geopolitical or national events that have privacy implications~\cite{westin2003social}. Recently, the COVID-19 pandemic created a global crisis. Governments around the world responded to this crisis by deploying new collection of personal data and new technologies. While intended to mitigate the spread of the COVID-19 virus, these new data uses have also been perceived as infringing on the privacy of citizens' personal data~\cite{ahmad2020state}. This raises the question of whether people's sentiments toward related use of their data -- i.e., for governmental or health purposes outside the context of the pandemic -- have changed.

People's privacy concerns and opinions influence digital privacy-related legislation, as well as their adoption of technologies and willingness to share their personal information~\cite{baruh2017online,citron2016privacy} -- albeit with known biases~\cite{acquisti2015privacy}. Thus, it is critical for technologists to understand people's sentiments toward various uses of their personal data to ensure that technology is built in ethical alignment with the broader population, and that these technologies will ultimately be used~\cite{lee:dighum_perspectives:2021,ahmad2020state}. It is important for us to understand such sentiments in general, as well as during and in response to times of crisis or urgency, where rushed deployment of potentially privacy-infringing technologies may be pushed to mitigate harms in the short-term but may cause long-term impacts to both privacy rights and sentiments~\cite{westin2003social,ahmad2020state,alshawi2022data}.

Understanding privacy sentiments and digital norms, as well as threats that certain technologies may pose, can inform the creation of technologies and policies with appropriate privacy considerations that pave the way for higher adoption rates and less privacy threats from the beginning~\cite{newlands2020innovation}. Most studies examine privacy sentiment at single points in time; however, it is also necessary to study privacy sentiment across time to gain a deeper, anticipatory understanding of how and why sentiment changes and evolves, which can in turn allow us to \textit{design for} rather than \textit{react to} people's privacy preferences.

In this work, we use repeated cross-sectional surveys conducted using a composition of panels to measure privacy sentiment over time, to understand how people in the U.S. have changed their views on privacy during the COVID-19 pandemic. Following contextual integrity (CI) theory~\cite{nissenbaum2004privacy}, which suggests that people's data privacy sentiments relate to the use of their data in specific \textit{contexts} composed of attributes such as the entities accessing data and the purposes for which data are used, we focus on changes in privacy sentiment related to two data use contexts that are closely related to the data uses introduced during the COVID-19 pandemic: use of data by the government for public safety and use of data for individual health. In so doing, we seek to answer the following research questions: 
\begin{enumerate}[start=1,label={\bfseries RQ\arabic*:},wide = 0pt, leftmargin = 3em]

    \item Did people's attitudes toward governments' data uses and/or health-related data uses change after the onset of the COVID-19 pandemic, from 2019 to 2020?
    \item Did any changes observed in RQ1 sustain or subside after the onset of the pandemic, between 2020 and 2021?
    \item Do changes observed in RQ1 and RQ2 differ across sociodemographics, particularly gender, age, race, ethnicity, education level, and/or political leaning?
    
\end{enumerate}
To answer these questions, we statistically analyze survey data collected in June 2019, May through June 2020, and June 2021 (total $n=6,676$). 

After the onset of the COVID-19 pandemic (\textbf{RQ1}), from 2019 to 2020, we observe a significant decrease in people's odds of accepting the use of their data for government assessment of terrorism threats, and significant increases in both the acceptance of fitness tracker data for medical research and the acceptance in use of social media data for detecting and intervening in mental health. A year later (\textbf{RQ2}), in 2021, these changes were sustained (i.e., no further significant changes among these items was recorded), and additionally, people were less accepting of law enforcement use of genetic data for crime solving when compared to their 2019 acceptance levels. 

We find also that sentiment changed within demographic groups primarily between 2019 and 2020 (\textbf{RQ3}). The observed socio-demographic changes led to increased consensus within different socio-demographic groups who previously had divergent attitudes regarding privacy. A notable exception is in the attitudes of Republicans vs. Democrats, which remained divergent but changed in direction in 2021 following the November 2020 presidential election. 

Drawing on our data and analysis, we discuss implications for research and design, including identifying potential predictors of future privacy sentiment shifts.
\section{Background and Related Work}

Here we provide an overview of the findings and impact of prior work examining privacy sentiment over time. Additionally, we review prior work on people's sentiments regarding data privacy in the contexts relevant to our work: the COVID-19 pandemic, government use, and health.

\subsection{Privacy Sentiment Over Time}
\label{sec:rel:overtime}
In the United States, views towards privacy have been known to shift over time. In the first nationally representative survey on different dimensions of privacy in the U.S., Westin~\cite{westin2003social} reported that in 1978 most respondents never felt like their privacy had been invaded, though 64\% also said that they were ``concerned'' over privacy threats, an increase from prior, smaller privacy-sentiment measurements, which he attributed to revelations of privacy invasion such as the Watergate scandal. The proportion of those in the U.S. with privacy concerns grew to 84\% by 1995~\cite{westin2003social}. Katz and Tassone~\cite{katz1990report} also found that privacy concern rose in the 1980s through the 90s, and that those in the U.S. speculated that privacy would become a larger problem in the future. Best et al.~\cite{best2006privacy} identified further growth in privacy concern from the 1990s to 2000, with growth in concern accelerating in tandem with growth in online privacy invasions beginning in the mid-1990's in the wake of increasing popularity of the internet.

Making sense of these changes over time can be aided by situating public opinion in the context of important social or political changes~\cite{westin2003social}. While they may not fully explain changes in public sentiment around privacy, impactful events, especially those with privacy implications, provide possible rationale for large scale public opinion shifts. One salient example is the terrorist attacks in New York City on September 11, 2001. Prior to September 11, U.S. polling data showed an increasing percentage of respondents who viewed government data gathering as a serious threat to privacy between 1985 through 1996~\cite{anthony2015big}, as well as an increasing percentage who were concerned about threats to their personal privacy in general between 1990 and 2000~\cite{best2006privacy}. Surveys taken after September 11 show a sharp increase in public support for giving up civil liberties and privacy protections for security against terrorism~\cite{best2006privacy}; a substantial increase in trust in government was also seen~\cite{chanley2002trust}. Following the immediate aftermath of September 11, a ``rebound'' effect was observed as support for giving up civil liberties and concern over government surveillance returned to similar levels as before the terrorist attacks~\cite{best2006privacy}.

Some prior work indicates that the approval of sharing data with the government has continued to decrease post-9/11. Possibly exacerbated by Edward Snowden’s leaking of classified National Security Agency information in 2013, over half of those surveyed disapproved of the U.S. government’s collection of phone and internet data for anti-terrorism purposes in the months following the leaks~\cite{geiger2018americans}. During this time, data also showed that U.S. respondents believed that the government was using their data for uses beyond combating terrorism, with more people expressing concern over protecting civil liberties than national security~\cite{dimock2013few}. People in the U.S. continued to disapprove of the monitoring of average American citizens in the years that followed~\cite{rainie2015americans,volz2017most}.

Prior literature~\cite{baruh2017online,citron2016privacy,westin2003social} suggests that public privacy sentiments play a substantive role in the enactment of privacy legislation and the governance of data use by both corporations and the government. In turn, such legislation informs the design of new technologies~\cite{phillips2004privacy}. Further, prior work suggests that privacy sentiments influence people's willingness to use such newly created technologies~\cite{marakhimov2017consumer}. Thus, gaining a large-scale and cultural-, time-specific understanding of privacy attitudes helps to anticipate shifts in technology policy, design, and use.

\subsection{Data Privacy and COVID-19}

In early 2020, the COVID-19 virus spread throughout the world, causing mass casualties and disruptions to daily life at an unprecedented scale \cite{NYTimesCovid}. Measures to control the spread of the virus were quickly implemented by governments across the globe, and for many of these solutions, the use and sharing of personal data was central. For instance, location and/or Bluetooth data was utilized for digital contact tracing, which helped health authorities trace the spread of COVID-19 and notify participating users when they came into contact with an infected person~\cite{ahmed2020survey,redmiles2020user}, and in some cases for mapping risk areas~\cite{desjardins2020rapid,budd2020digital}. Health data like body temperatures and negative COVID-19 test results were provided in exchange for permission to enter buildings or travel~\cite{Repko01,Schwartz01,Begley01}. Non-health related data like online shopping information and credit card transactions were utilized to track possible virus exposures \cite{ahmad2020state}. Some countries even deployed public surveillance measures to enforce stay-at-home policies via location-tracking wearables and drones~\cite{couch2020covid}.

Research has found that these data uses have varying levels of public approval, with participants voicing privacy concerns across studies (e.g.,~\cite{zhang2020americans,simko2020covid,kaptchuk2020good,williams2021public,kim2021examination,redmiles2020user}). For instance, Altmann et al. \cite{altmann2020acceptability} find that COVID-19 contact tracing apps garnered high approval across multiple countries, yet identified concerns about privacy and a lack of trust in government being two key variables that hinder adoption. In contrast, Utz et al. \cite{utz2021apps} find that acceptance of contact tracing apps and data sharing differed by country and were dependent on the country's data sharing norms; moreover, Trang et al. \cite{trang2020one} explored sentiments in Germany and found that social benefits and convenience are more important factors than privacy for citizens who were undecided or against contact tracing apps. Seberger and Patil ~\cite{seberger2021post} find that with contact tracing, participants identify a trade-off between adopting technology that would help society for the "greater good," while also infringing on their privacy and sharing their data with third parties -- importantly, the idea of the greater good also seemed to have a "shelf life," and was most important to participants in the context of a pandemic.

Most related to our work, Biddle et al. \cite{biddle2022data} finds that over the course of the pandemic, respondents in Australia grew less concerned about the use of their personal data by organizations broadly, compared to their sentiments before the pandemic. We take a similar approach and work within the context of the United States to investigate whether the extended use of personal data during the pandemic may have impacted attitudes beyond those directly relating to COVID-19 mitigation technology. Namely, as the pandemic is a widespread medical phenomena which governments around the globe played key roles in sharing information about and deploying mitigation technologies for, we hypothesize that broader attitudes, outside specific pandemic contexts, regarding 1) sharing data with governments and authorities for public safety purposes, and 2) sharing health-related data and/or data for health-related purposes, may have changed since the onset of the pandemic. We explore how the views of those in the U.S. on data privacy have changed in these contexts.

\subsection{Government Data Use}
\label{sec:relwork:government}

Our research builds on the body of work documenting shifts in public opinion on government use of personal information, with a focus on how attitudes may have changed throughout the COVID-19 pandemic. Government surveillance involves federal entities collecting data on civilians, often to exert social control~\cite{fuchs2017internet}. As aforementioned, Westin~\cite{westin2003social} contends that public trust in government and willingness to accept their use of civilian data has risen and fallen throughout history in response to political happenings such as 9/11 or privacy scandals like Watergate.

In the context of the pandemic, Zhang et al.~\cite{zhang2020americans} find modest to low levels of support for government surveillance measures to combat the spread of COVID-19 in the U.S., like encouraging use of contact tracing apps or implementing immunity pass systems. Additionally, Simko et al.~\cite{simko2020covid} find that while participants acknowledge that the benefits of data sharing with the government during the pandemic outweigh the risks, a majority of participants doubted that the government would delete their data or use it solely for COVID-19 related purposes. Our research expands this work and investigates how views on government uses of data that are not directly related to COVID-19 relief efforts may have changed during the pandemic.

More broadly, in recent decades, Western societies have increasingly used and relied on ``surveillance-oriented security technologies'' to proactively combat terrorism and other crimes~\cite{pavone2012public}; and, as mentioned, measures to prevent the spread of COVID-19 like contact tracing apps have made government use of civilian data all the more salient. Turow et al.~\cite{turow2018divided} captures the complexity in attitudes towards ``everyday surveillance practices,'' finding that government and law enforcement surveillance practices evoked the most emotional division in participants, with similar percentages of those in the U.S. saying they feel happy vs. sad, and pleased vs. angry, that they take place. Such complexity may explain divergence in existing measurements of sentiment toward government data collection. Some recent measurements suggest that a majority of those in the U.S. disapprove of government surveillance for terrorism and other public safety purposes~\cite{auxier2019americans}. Other work suggests that the government’s use of big data and digital surveillance technologies is becoming normalized in modern society, largely through its necessity for various administrative functions and perceived benefits like crime control~\cite{hu2017national}. Rather than directly opposing surveillance, citizens may resign them selves to acceptance based on their own perceived lack of knowledge and control~\cite{dencik2017advent}.

We examine how opinions on governments' and authorities' use of personal data for the purposes of security and safety have changed across time in tangent with social and political circumstance, and use two specific scenarios to explore this question as described in  Section~\ref{sec:meth}.

\subsection{Privacy and Health-Related Data}
\label{sec:relwork:health}
Broadly, prior work finds that personal health information is ``very sensitive'' for most people in the U.S.~\cite{madden2014public}. When asked about concerns towards health data sharing, studies with international populations find concerns around a lack of control over their data, such as misuse or overuse of their data beyond the use to which they consented, and concerns related to a lack of anonymity~\cite{howe2018systematic,hill2013let,aitken2016public,cheung2016privacy,kalkman2019patients,stockdale2018giving,richter2019patient}. Yet, despite these concerns, Howe et al.~\cite{howe2018systematic} find that across health-data-sharing research studies, participants are willing to share their data for medical benefits at the individual and societal level. For instance, people in the U.S. are accepting of online or electronic health systems that improve the patient care experience~\cite{rainie2016privacy,gaylin2011public}. Additionally, data from international populations also show the concept of ``the greater good,'' referring to advances in the medical field that would benefit the general public, motivates people to allow use and sharing of their data~\cite{aitken2016public,stockdale2018giving,spencer2016patient}.

Research from Europe shows people generally approve of sharing their data with healthcare professionals~\cite{darquy2016patient,haddow2011nothing}, while research from the U.S. shows that university researchers within the U.S. and relevant non-profit organizations are also generally acceptable sources ~\cite{majumder2016beyond,goodman2017identified}. However, work from the U.S. and beyond also finds that people are least willing to share their health data with private companies or other profit-seeking ventures~\cite{hill2013let,aitken2016public,goodman2017identified}. Among those who are willing to consider sharing their data with for-profit entities, Trinidad, Platt and Kardia~\cite{trinidad2020public} find that respondents are more comfortable with companies accessing their health data for purposes related to their care as a patient, compared to business purposes.

The variation and nuance in opinions on sharing general health-related data are also present in COVID-19 specific literature, where themes of privacy, consent, medical benefits and social good tend to be at odds with pandemic mitigation technology \cite{williams2021public,geber2022tracing,walrave2020adoption,kim2021examination}. 
The nature of the COVID-19 pandemic as a public health crisis, the importance of health-related data to COVID-19 relief efforts, and the increased digitization of healthcare during the pandemic~\cite{koonin2020trends}, lead us to ask in our work whether the onset of the pandemic impacted sentiments toward health-related data privacy.

\section{Methodology}
\label{sec:meth}

We use repeated cross-sectional surveys to investigate changes in privacy sentiment among people in the U.S. We build on existing work examining privacy views across time, particularly in the context of significant events, and explore how data privacy sentiments of those in the U.S. have changed from 2019 to 2021, a period filled with salient geopolitical and national changes; chiefly, the COVID-19 pandemic, but also national events like the 2020 Presidential election. We focus our investigation on changes in privacy sentiment in two contexts~\cite{nissenbaum2004privacy} that are closely related to the context of data uses in the COVID-19 pandemic: the use of data by the government (or sub-entities) for public safety and health-related data and data use.

\subsection{Scenarios to Measure Privacy Sentiment} 

We measure changes in privacy sentiment related to government and health contexts using four specific scenarios. The use of specific scenarios to extrapolate sentiments toward more general constructs is in line with epistemological studies (e.g. \cite{kirk1986reliability}) showing that concrete scenarios produce more accurate measurements of sentiment, particularly for hard-to-measure constructs such as privacy sentiments~\cite{preibusch2013guide}. For each specific scenario, we ask respondents: "In your opinion, do you think the following uses of data or information by the government or private companies are acceptable or unacceptable?" with the answer choices "Acceptable," "Unacceptable," or "Unsure" to choose from. Below, we enumerate our four specific scenarios and explain their relevance in both the context of the COVID-19 pandemic and data privacy literature more broadly.

\medskip
\noindent \textit{Item 1: "The government collecting data about all Americans to assess who might be a potential terrorist threat."}

Our first government data use scenario focuses on broad collections of data by the government for combating terrorism. As summarized in Section~\ref{sec:rel:overtime}, prior work finds that people's privacy sentiments toward government data use in the U.S. have often shifted in response to national or international safety-related events such as 9/11 or changes in awareness of government data use. Prior research finds contradicting results regarding whether government data use is becoming more or less acceptable, underscoring the complexities of sentiments toward this topic in the U.S.

In urgent situations or crises, such as the COVID-19 pandemic, citizens may be more willing to share their data with necessary government organizations and authorities. For instance, in 2016, when the FBI ordered technology company Apple to bypass their encryption systems to unlock the iPhone of a mass shooter, Pew Research~\cite{pew2016more} found that just over half of respondents sided with the FBI in this case. Additionally, the propensity to share data with authorities, doctors and scientists for the ``greater good'' has been found to be more salient during the COVID-19 pandemic than outside of that context \cite{seberger2021post}.

Thus, to update our understanding of sentiment toward government data use for public safety purposes, particularly in the context of an international public safety (but not terrorism-related) event, we explore whether the attitudes of those in the U.S. towards this particular form of government data use changed over the course of the COVID-19 pandemic.

\medskip
\noindent \textit{Item 2: ``DNA testing companies sharing their customers’ genetic data with law enforcement agencies in order to help solve crimes.''}

Along with exploring sentiments related to ``the government'' using generalized "data" from those in the U.S., we explore changes in sentiment related to the use of specific types of health data (i.e., genetic data) for public safety by specific entities (i.e., law enforcement). This approach aligns with CI theory~\cite{crasnowFeministStandpointTheory}, which suggests that people's data privacy sentiments also relate to the use of specific data by specific entities.

Recent advances in genetic science have unlocked many uses for DNA data in fields like ancestry tracing, disease research, and criminal justice~\cite{Greguska01,doj01}. These uses have inspired conversations around the ethics of collecting and sharing genetic data. People perceive risks to sharing biometric data like genetic data -- such as privacy breaches and unauthorized uses -- along with benefits, like aiding research and societal welfare~\cite{shabani2014attitudes,lemke2010public}. Prior work in the U.S. and Canada show that people are generally willing to share genetic data with researchers or with research databases and biobanks~\cite{sanderson2017public,kaufman2009public,rogith2014attitudes,joly2015fair}. Kaufman et al. find differences in acceptability of various stakeholders accessing personal genetic data, with academic researchers being most acceptable~\cite{kaufman2009public}.

Genetic data is increasingly being used by law enforcement as genealogy companies become more popular with consumers and accessible online genetic databases grow ~\cite{skeva2020review}. For example, in 2018 police used the genealogy website GEDMatch to identify and arrest a 72 year-old man on suspicion of being the ``Golden State Killer,'' a criminal accused of committing high-profile murders throughout the 1970s and 80s, due to his matching DNA~\cite{Kolata01}. Since then law enforcement have been able to revisit over 50 cold cases with new leads acquired from genealogy databases~\cite{kennett2019using}. This method of crime solving is new, leaving it largely unregulated~\cite{hazel2020world}. One study suggests that a majority of those in the U.S. support law enforcement use of genealogy websites~\cite{guerrini2018should}. Our work hopes to gauge current public opinion on the practice, particularly in the wake of expanded government use of health data for public safety and the growing discourse surrounding the role of police and law enforcement in communities, which increased significantly~\cite{nytblm} during the COVID-19 pandemic~\cite{patnaude2021public,bolsover2020black,jones2020potential,jennings2020immediate,Pew01}.

\medskip
\noindent \textit{Item 3: ``Makers of a fitness tracking app sharing their users’ data with medical researchers seeking to better understand the link between exercise and heart disease.''}

In addition to government-related uses of data, we investigate sentiments towards health-related data uses, including both use of health-related data (as addressed in this item) and use of non-health data for health-related purposes (as addressed in Item 4, below). As detailed in Section~\ref{sec:relwork:health}, people consider the benefits of sharing their health-related data when making their decision to do so. Our first health-related scenario asks specifically about sharing of data from wearable fitness tracking technology, an increasingly popular technology~\cite{thompson2018worldwide,piwek2016rise}. Data generated by these technologies can include anything from users’ heart rates, to sleep patterns, to number of steps~\cite{fietkiewicz2020fitness}. A number of privacy-related factors including trust in the device~\cite{pfeiffer2016quantify} and the reputation of the company producing it~\cite{adebesin2020mediating}, as well as perceived privacy risks~\cite{gao2015empirical,li2016examining} are all highly influential in user adoption of such trackers.

Relevant to our study, Wiesner et al.~\cite{wiesner2018technology} find that while most of the German fitness tracker users they surveyed were unconcerned with their data being shared without their consent, only one in seven were willing to actively share their data for research purposes. In light of the COVID-19 pandemic, when the use and sharing of health-related data for research has become increasingly necessary and urgent, we investigate whether views on sharing fitness data for research have shifted.

\medskip
\noindent \textit{Item 4: ``A social media company monitoring its users’ posts for signs of depression, so they can identify people who are at risk of self-harm and connect them to counseling services.''}

Finally, we ask respondents in our survey how accepting they are of using social media data for the purpose of identifying mental health crises and providing services. Increasingly, technologies that can recognize emotions and moods using various data like biometrics and online behavior are being developed and deployed~\cite{andalibi2020human}. In the context of social media, researchers have analyzed data from users’ posts, online activities, and profile characteristics on platforms like Twitter, Facebook, and Reddit to predict mental health issues in users ~\cite{tsugawa2015recognizing,shen2017depression,hussain2020exploring,katchapakirin2018facebook,tadesse2019detection,de2012happy,de2014characterizing,reece2017instagram}. Insights can be derived at the population level ~\cite{de2013social,schwartz2014towards} or at the individual level ~\cite{o2018rate,coppersmith2018natural}. Social media platform Facebook has disclosed their active use of both human and artificial intelligence to identify users who may be in mental health crises based on their posts, in order to get first responders and resources to them~\cite{FB01,FB02}.

Analyzing social media content for mental health information has benefits, the most prominent being the opportunity to provide mental health care to at-risk individuals. But prior work has also noted concerns related to privacy and ethics -- for instance, the difficulties of gaining fully informed consent from users whose social media data are monitored for mental health purposes~\cite{nicholas2020ethics}, or the unresolved question of whether social media posts are public or private McKee~\cite{mckee2013ethical}. Chancellor et al.~\cite{chancellor2019taxonomy}, in their taxonomy of existing methods for deriving mental health information from users’ social media posts, acknowledge the benefits of early detection of mental disorders while also noting the risks to users such as inaccurate mental health predictions and a lack of proper data protections.

The limited body of research on how users feel about this practice has produced, like other studies relating to opinions on health data uses, complex results. Andalibi and Buss find that the concept of emotion detection and prediction on social media evoked feelings of discomfort in social media users~\cite{andalibi2020human}. Meanwhile, a U.S. focus group study revealed that there were mostly favorable views towards using Twitter data to monitor mental health at a population level~\cite{mikal2016ethical}, while a survey of people in the UK found that most social media users supported data analysis on Facebook content for the purpose of targeting mental health resources, but less than half were willing to give consent for their own Facebook data to be analyzed this way~\cite{ford2019public}. Our research builds on this body of research and investigates changes in sentiment during the COVID-19 pandemic, during which mental health challenges have been an increasingly important public health topic~\cite{KFF01,CDC01}.

\begin{table*}[t]
\centering
\footnotesize
\begin{tabular}{c|c|c|c|c|c|c}
    \toprule
&{\bf Metric (\%)} &{\bf 2019 [Q1,Q2]} & {\bf 2019 [Q3,Q4]} &{\bf 2020} & {\bf 2021} & {\bf 2020 U.S. Census}\\
\midrule
\parbox[t]{2mm}{\multirow{2}{*}{\rotatebox[origin=c]{90}{Gender$^*$}}}&Man & 44 & 45 & 56&47 &47\\
&Woman & 56 & 55 & 44 &53 & 50\\
& Non-binary & -- & -- & 0.4 & 1 & 3\\
\midrule
\parbox[t]{2mm}{\multirow{5}{*}{\rotatebox[origin=c]{90}{Race/Ethn.}}}&White & 78 & 78 &48 &65& 62 \\
&Hispanic & 14 & 13 &19 &10&19 \\
&Black/African American & 12 & 11& 22 &16 &14 \\
&Asian/Asian American & 3 & 3 &12& 10& 7 \\
&Other & 8 & 8 &19& 10& 10 \\
\midrule
\parbox[t]{2mm}{\multirow{3}{*}{\rotatebox[origin=c]{90}{Educat.}}}&H.S. or Less & 34 &34& 30 &25&38\\
&Some college & 28& 28 &30 &35&26\\
&B.A./B.S. or above &39 & 38 &40 &40&35\\
\midrule
\parbox[t]{2mm}{\multirow{4}{*}{\rotatebox[origin=c]{90}{Age}}}& 18-29 years & 15 & 15 & 21& 24&16\\
&30-49 years & 31 & 30 &39& 40 &26\\ 
& 50-64 years & 30 &31 & 23 & 18 &19\\
& 65+ years & 23 &23 & 17 & 18 &16\\
\midrule
\parbox[t]{2mm}{\multirow{2}{*}{\rotatebox[origin=c]{90}{Pol}}} & Dem/Lean Dem & 56 & 55&64&63&--\\
&Rep/Lean Rep & 45 & 45&36& 37&--\\
\bottomrule
\end{tabular}
\caption{Demographics for our four samples as compared to the demographics of the U.S.~\cite{Educatio77:online,Improved82:online,National8:online}. $ ^*$As noted in Section~\ref{sec:meth:survey} Pew uses a binary gender question, while our survey measures a broader range of gender identity in line with best practice~\cite{spiel2019better}.} 
\vspace{-2ex}
\label{tab:demo}
\end{table*}
\subsection{Survey Methodology}
\label{sec:meth:survey}

We use the aforementioned items to assess sentiment toward government and health data privacy in 2019, 2020, and 2021. We use a repeated cross-sectional survey methodology. That is, we collect responses from different respondents in the same population (the U.S.) each year via online surveys. Repeated cross-sectional surveys may be conducted using the same panel in every year or using a composition of panels across years.\footnote{Multiple prior works use a composition of panels. Some statistically compare data across different panels owned by different companies (e.g., \cite{regan2013generational}) while others (e.g., \cite{hammond2019prevalence}) compare data collected by a single company such as Nielsen, one of the largest consumer polling companies in the U.S., which leverages a combination of different panels with different sampling methodologies in each year.} We use the latter approach in our work: we draw our data from one survey panel in 2019 and from another in 2020 and 2021.

Specifically, the 2019 privacy sentiment dataset that we use in our analysis was collected by Pew Research Center (Pew) in June 2019 via their American Trends Panel~\cite{auxier2019americans}. 
Pew showed the two government data items to a subset of their panelists ($n=2,012$) and the two health data items to a different subset of their panelists ($n=1,989$).

The 2020 ($n=1,138$) and 2021 ($n=1,537$) datasets we use were sampled from the Cint survey panel. In May through June 2020, and in June 2021, we administered the same four items using the exact same phrasing, as well as a series of demographic questions assessing the respondents' age, gender identity, race, ethnicity, level of educational attainment, and political leaning. We administered both surveys online, as was done in the original Pew survey.

\medskip
\noindent\textbf{Sampling.}
Recruitment for the Pew survey panel occurs in a two step process: first they recruit a population of roughly 10,000 individuals who agree to take their surveys, and then for individual surveys, they sample from that population. Recruitment for the 10,000 panelists is done via mailers and then individual surveys are (primarily) taken online.
\footnote{For more detail on the panel methodology, see \url{https://www.pewresearch.org/our-methods/u-s-surveys/the-american-trends-panel/}.} 
The dataset we use is the Wave 49 American Trends Panel, which focuses on privacy and surveillance. Since these data were collected as a sub-sample of the overall Pew panel, we see deviance from the 2020 Census demographics in this panel. Specifically, while Pew aims to provide very representative data, and their data have become an industry standard that are heavily leveraged in academia and legislation, we note that in the 2019 Pew survey there was an over-representation of white individuals. We discuss how our analysis procedure addresses this over-representation in Section~\ref{sec:meth:analysis}, below.

Cint recruits for their survey panel in a similar manner to Pew. They recruit a large group of panelists via mailers, frequent flyer programs, online advertisements, and other approaches. For each survey, they sample a subpopulation of the larger panel. Respondents to the surveys we sampled via Cint's panel took these surveys online, like the Pew participants, and we ensured that those who took the survey in 2020 did not take it again in 2021. We also requested that the demographics of our respondents be representative of the U.S. population in terms of age, gender, education, race, and ethnicity; specifically, we set quotas that the proportion of respondents in each category would be within $\pm$10\% of the U.S. Census. In one instance, this quota was not met: our 2020 Cint panel was under-representative of white individuals. However, as discussed further in Section~\ref{sec:meth:analysis}, we still had sufficient statistical power for white individuals and used specific analysis methodologies to account for demographic variance between the samples collected in each year and ensure that these variances and do not compromise the integrity of our results and interpretation.

\medskip
\noindent\textbf{Demographics.}
Demographic variables are given by self-report data from individual respondents in each survey panel. In 2019, Pew captured and reported: race and ethnicity, education level, age, and political leaning. We captured the same variables in the surveys we ran in 2020 and 2021.
Pew also captures a variable they term `sex' for which they have only binary responses. In our surveys, we asked about gender in an inclusive manner following~\citep{spiel2019better}.

Our analysis uses the following demographic variables: gender (women vs. men \footnote{We received 5 non-binary responses in 2020 and 15 in 2021. As we lack statistical power to draw meaningful conclusions about the privacy sentiments of the non-binary community in specific, we do not include these data in our analysis. However, this in no way diminishes the importance of understanding the unique and valuable insights of this population, which should be explored with either larger samples (yielding statistical power) or targeted research specific to their viewpoints and perspectives~\cite{geeng2018queer,geengqueer22}.}), age (as a categorical variable), race (as a categorical variable), whether the participant was Hispanic, education (as a categorical variable), and political leaning (as a binary variable). The categories for each variable can be found in Table~\ref{tab:demo}.

\medskip
\noindent\textbf{Ethics.} The methods for our data collection (2020 and 2021 datasets) were approved by our institution's review board. The 2019 data were collected by Pew. We use only the de-identified data that Pew publicly released on their website in our analysis. The full Pew data report can be found in~\cite{auxier2019americans} and an overview of their ethical principals can be found at: \url{https://www.pewresearch.org/about/our-mission/}.

\subsection{Analysis}
\label{sec:meth:analysis}
Our primary outcome of interest (dependent variable) was whether a respondent found the data use presented in each of the four items acceptable. We built three sets of logistic regression models to understand sentiment toward these data uses and particularly changes in that sentiment.

We use regression as our primary method for analysis because it accommodates for demographic differences between our three datasets. We test for significance of odds ratios of learned parameters in logistic regression models. These tests do not assume equal counts of different demographic groups~\citep{oddsratio}. The use of this analysis technique allows us to rigorously and appropriately analyze our data, despite a lack of consistent demographic apportionment between the Census, Pew, and Cint.

To address RQ1 and RQ2, we report on overall changes between each pair of years (2019-2020, 2020-2021, and 2019-2021) by using regressions which include the year as a linear predictor and control for demographic effects; these results can be found in Table~\ref{tbl:year}. In these regressions, an odds ratio below 1 would indicate a \emph{decrease} of acceptability in the second year compared to the first, and an odds ratio above 1 would indicate an \emph{increase}. 

In addressing RQ3, we first report on baseline effects using our regression data rather than using raw numbers in order to control for demographic variance between our various samples. As such, we construct a separate logistic regression model for each item and for each year that predicts the dependent variable from the demographic information collected in the surveys. Further information on the categories for each variable can be found in Table~\ref{tab:demo}. This second set of regressions tells us which sociodemographics were significant for a given question in a given year, but do not allow us to reason about changes in sociodemographics between years. Therefore, we construct models for each pair of years (2019-2020, 2020-2021, and 2019-2021) to evaluate whether the demographic changes are sustained. In these models, we treat the year as an interaction term with the demographic variables, which allows us to statistically test whether an observed demographic change from one year to another was significant or just occurred by chance. The relevant results of these regressions are shown in Figure~\ref{tab:demo}.

The regression models provide significance values for odds ratios found from the regression's fitted values. Significance values for our conclusions were set with $\alpha=0.05$. See~\cite{oddsratio} for a deeper explanation of odds ratios and logistic regressions.

Finally, we note that, consistent with the literature \cite{gelman2012we}, because our data are organized from three similar, non-identical populations and the comparisons we make are purely complementary, we do not make any multiple comparison corrections in our regressions. 

\subsection{Limitations}
\label{sec:meth:limitations}
The data used in our work comes from respondents who were sampled from two different panels. While this approach of comparing across multiple panels or drawing from data composed from multiple panels is extant in prior work (see e.g., \cite{regan2013generational,hammond2019prevalence}), we acknowledge that the difference in Pew’s sampling methodology vs. our own may be confounding when determining attitude changes across time.  However, both samples used the same mode -- online survey -- as well as identically phrased questions of interest. Further, as described in Section~\ref{sec:meth:analysis}, to mitigate the effects of demographic variance between our samples we report all baseline and change effects in the context of logistic regression models that control for demographics.

Additionally, while our interest is in how privacy sentiment changed over the course of the COVID-19 pandemic, our analysis is not causal. We report on changes observed between 2019, 2020, and 2021 but cannot definitively conclude that particular events (like the onset of the pandemic) caused these changes.

Finally, our work is subject to limitations typical to most survey work: respondents may have been vulnerable to social-desirability bias or the privacy-specific ``privacy paradox,'' and as a result may have reported stronger or different opinions than those they actually held. To mitigate the former, Pew engages in extensive pre-testing and multiple rounds of question drafting, editing, and piloting as described in more depth at \url{https://www.pewresearch.org/our-methods/u-s-surveys/writing-survey-questions/}. Regarding the latter, we note that we study privacy sentiments rather than privacy behavior; the latter is the chiefly effected variable of the privacy paradox. While privacy sentiment does not always accurately predict digital behavior, these sentiments still influence policy makers and have been shown to be among multiple behavioral factors related to people's technology use~\cite{acquisti2015privacy,baruh2017online,citron2016privacy}; thus, they still merit study, especially over time.
\section{Results}
\begin{figure*}
    \centering
    \includegraphics[width=.9\linewidth]{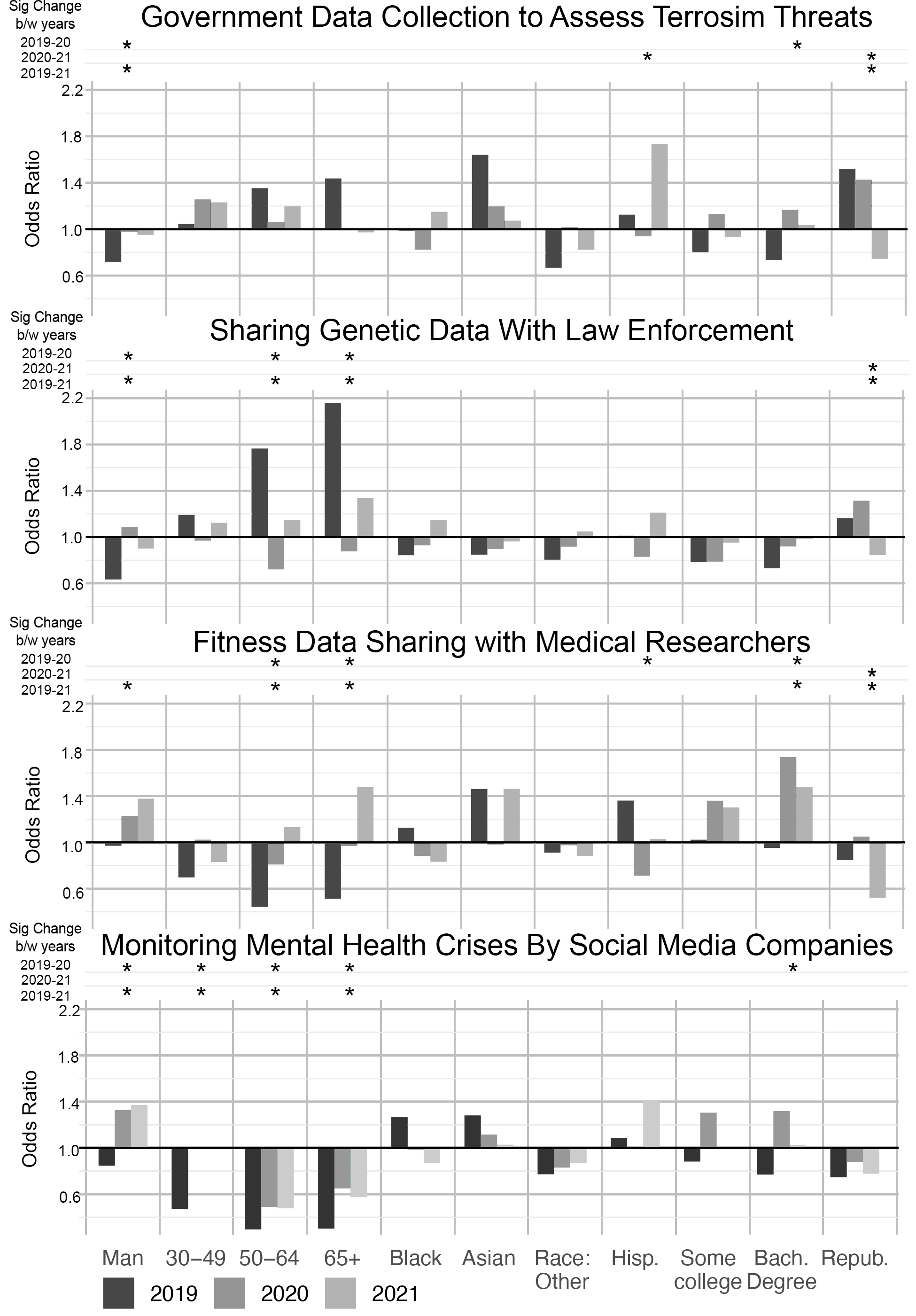}
    \caption{Summary figure reporting odds ratios from regressions for each survey item and each year. Bars depict odds ratios from intra- year regressions. At the top of each chart, we use stars to summarize significant \textit{inter}-year differences for the different demographic groups. Table~\ref{tbl:intra_years} details the results of the intra-year models and Table~\ref{tbl:inter_years} the inter-year models.}
    \label{fig:summary_results}
\end{figure*}

We analyze the results from each item individually. Results\footnote{For reference, analysis code and all regression tables can be found here in Appendix Tables~\ref{tbl:QuestionA_years} - \ref{tbl:model_2019_2021}} are summarized in Figure~\ref{fig:summary_results}.  
Regressions supporting RQ1 can be found in Table~\ref{tbl:year}. Regressions supporting RQ2 can be found in Table~\ref{tbl:intra_years}. Regressions supporting RQ3 can be found in Table~\ref{tbl:inter_years}.

\begin{table*}[h!t] \centering 
\caption{Overall changes between in perceived acceptability between each pair of years with odds ratios, confidence intervals, and $p$-values even  when controlling for demographic effects.
}
  \label{tbl:year} 
  \small
\begin{tabular}{@{\extracolsep{5pt}}lcccc} 
\\[-1.8ex]\hline 
\hline \\[-1.8ex] 
 & Item 1 & Item 2 & Item 3 & Item 4 \\ 
 & Government/Terrorism & Law Enforcement/Genetic & Medical Research/Fitness & Corporate/Mental Health \\ 
\hline \\[-1.8ex] 
 2019-2020 & 0.782 & 0.942 & 1.264 & 2.180 \\ 
  & (0.668, 0.915) & (0.805, 1.102) & (1.080, 1.479) & (1.846, 2.574) \\ 
  & p = 0.003$^{**}$ & p = 0.456 & p = 0.004$^{**}$ & p $<$ 0.001$^{**}$ \\ 
  & & & & \\ 
 2020-2021 & 0.908 & 0.897 & 0.987 & 0.948 \\ 
  & (0.773, 1.066) & (0.765, 1.052) & (0.841, 1.159) & (0.806, 1.114) \\ 
  & p = 0.239 & p = 0.183 & p = 0.874 & p = 0.514 \\ 
  & & & & \\ 
 2019-2021 & 0.680 & 0.814 & 1.246 & 2.039 \\ 
  & (0.591, 0.782) & (0.709, 0.936) & (1.083, 1.434) & (1.757, 2.365) \\ 
  & p $<$ 0.001$^{**}$ & p = 0.004$^{**}$ & p = 0.003$^{**}$ & p $<$ 0.001$^{**}$ \\ 
  & & & & \\ 
 \hline 
\hline \\[-1.8ex] 
\textit{Note:}  & \multicolumn{4}{r}{$^{*}$p $<$ 0.05; $^{**}$p $<$ 0.01} \\ 
\end{tabular} 
\end{table*}

\subsection{Government Data Collection\\to Assess Terrorism Threats}
Overall, we see that respondents acceptance of the government collecting data to assess a terrorism threat waned from 2019 into 2020 and stayed at a lower level in 2021 compared to 2019, even when controlling for demographic changes.

Considering 2019 as our baseline, we observe the following statistically significant effects. Men have significantly lower odds\footnote{In all cases where we describe a change throughout the paper, we refer to a change in the odds of acceptability, even if, for brevity, this phrase is not specifically used.} than women of finding terrorist assessments by the government acceptable. We also observe an increasing trend in acceptability by age, with the oldest two age groups rising to the level of statistical significance. We see that those who have a Bachelor's degree or higher have significantly lower odds of finding terrorist assessments by the government acceptable when compared to those with a high school degree or less. Finally, Republican or Republican-leaning individuals have significantly higher odds of accepting government terrorist assessments compared to their Democratic counterparts.

In 2020, however, we observe only one demographic variable that has significant differences when compared to its reference group, political party: Republicans have higher odds of finding government data use for terrorist assessment acceptable than Democrats.

The lack of other demographic differences in 2020 is due in part to statistically significant increases in acceptability among those with at least a Bachelor's degree and men.
When we examine whether these changes persisted into 2021, we see that men's decreased acceptance of this data use held in 2021. The Bachelors degree difference between 2019 and 2021 is just barely no longer significant, although the 2021 level is also not significantly different from the decreased 2020 level.

As for changes that were brought about in 2021, we see two significant differences: Hispanic ethnicity and political party. We observe that Hispanics became significantly more likely to find terrorist assessments by the government acceptable. Acceptability of terrorist assessments by the government is significantly moderated by a participant's political party affiliation both between 2019 and 2021 and between 2020 and 2021; specifically, Republicans, who were \emph{more} likely than Democrats to accept government data use for terrorist assessments in 2019 and 2020 were \emph{less} likely to accept it in 2021.

\begin{table*}[!htbp] \centering 
  \caption{Intra-year logistic regression models for the relationship between acceptance of a particular data use scenario and sociodemographics. Odds ratios are shown and * depict significance of relationship. For example, the first column depicts the Government/Terrorism question in 2019 where men were less likely than women (odds ratio less than one) to agree with the data use in 2019, people who were 50-64 and 65+ were more likely than those 18-29, etc.} 
  \label{tbl:intra_years} 
\resizebox{\linewidth}{!}{
\begin{tabular}{@{\extracolsep{5pt}}lcccccccccccc} 
\\[-1.8ex]\hline 
\hline \\[-1.8ex] 
 & \multicolumn{3}{c}{\textit{Item 1:}} & \multicolumn{3}{c}{\textit{Item 2:}} & \multicolumn{3}{c}{\textit{Item 3:}} & \multicolumn{3}{c}{\textit{Item 4:}} \\ 
 & \multicolumn{3}{c}{\textit{Government/Terrorism}} & \multicolumn{3}{c}{\textit{Law Enforcement/Genetic}} & \multicolumn{3}{c}{\textit{Medical Research/Fitness}} & \multicolumn{3}{c}{\textit{Corporate/Mental Health}}  \\ 
\cline{2-4} \cline{5-7} \cline{8-10} \cline{11-13} \\

 & 2019 & 2020 & 2021 & 2019 & 2020 & 2021 & 2019 & 2020 & 2021 & 2019 & 2020 & 2021 \\ 
\hline \\[-1.8ex] 
 Gender: Man & 0.717$^{**}$ & 0.977 & 1.369$^{**}$ & 0.633$^{**}$ & 1.087 & 0.900 & 0.971 & 1.228 & 1.377$^{**}$ & 0.848 & 1.326$^{*}$ & 1.369$^{**}$ \\ 
  & & & & & & & & & & & & \\ 
 Age: 30-49 & 1.045 & 1.257 & 1.007 & 1.191 & 0.969 & 1.124 & 0.697$^{*}$ & 1.023 & 0.831 & 0.473$^{**}$ & 1.002 & 1.007 \\ 
  & & & & & & & & & & & & \\ 
 Age: 50-64 & 1.353$^{*}$ & 1.062 & 0.480$^{**}$ & 1.765$^{**}$ & 0.721 & 1.147 & 0.442$^{**}$ & 0.809 & 1.134 & 0.297$^{**}$ & 0.490$^{**}$ & 0.480$^{**}$ \\ 
  & & & & & & & & & & & & \\ 
 Age: 65+ & 1.437$^{*}$ & 0.998 & 0.576$^{**}$ & 2.157$^{**}$ & 0.876 & 1.337 & 0.513$^{**}$ & 0.969 & 1.477$^{*}$ & 0.304$^{**}$ & 0.652$^{**}$ & 0.576$^{**}$ \\ 
  & & & & & & & & & & & & \\ 
 Race: Black & 0.986 & 0.823 & 0.871 & 0.843 & 0.928 & 1.149 & 1.127 & 0.882 & 0.833 & 1.264 & 0.987 & 0.871 \\ 
  & & & & & & & & & & & & \\ 
 Race: Asian & 1.640 & 1.195 & 1.026 & 0.847 & 0.897 & 0.962 & 1.461 & 0.982 & 1.463$^{*}$ & 1.280 & 1.114 & 1.026 \\ 
  & & & & & & & & & & & & \\ 
 Race: Other & 0.667$^{*}$ & 1.017 & 0.870 & 0.805 & 0.917 & 1.049 & 0.912 & 0.975 & 0.884 & 0.774 & 0.831 & 0.870 \\ 
  & & & & & & & & & & & & \\ 
 Ethn.: Hispanic & 1.124 & 0.941 & 1.413 & 1.012 & 0.829 & 1.211 & 1.361$^{*}$ & 0.713 & 1.029 & 1.085 & 0.992 & 1.413 \\ 
 & & & & & & & & & & & & \\ 
 Ed. Att.: Some college & 0.801 & 1.131 & 1.006 & 0.784$^{*}$ & 0.788 & 0.951 & 1.023 & 1.360 & 1.302 & 0.884 & 1.303 & 1.006 \\ 
  & & & & & & & & & & & & \\ 
 Ed. Att.: Bachelor's+ & 0.735$^{**}$ & 1.166 & 1.023 & 0.730$^{**}$ & 0.919 & 0.984 & 0.952 & 1.738$^{**}$ & 1.481$^{**}$ & 0.772$^{*}$ & 1.317 & 1.023 \\ 
  & & & & & & & & & & & & \\ 
 Political Id.: Rep./Lean Rep. & 1.518$^{**}$ & 1.427$^{*}$ & 0.779$^{*}$ & 1.164 & 1.314$^{*}$ & 0.844 & 0.848 & 1.050 & 0.522$^{**}$ & 0.748$^{*}$ & 0.880 & 0.779$^{*}$ \\ 
  & & & & & & & & & & & & \\ 
 Intercept & 1.014 & 0.571$^{**}$ & 0.942 & 1.006 & 1.006 & 0.754 & 1.191 & 0.687 & 0.778 & 1.099 & 0.802 & 0.942 \\ 
  & & & & & & & & & & & & \\ 
\hline 
\hline \\[-1.8ex] 
\textit{Note:}  &&&&&&&&&\multicolumn{4}{r}{$^{*}$p $<$ 0.05; $^{**}$p $<$ 0.01} \\ 
\end{tabular} }
\end{table*}

\subsection{Sharing Genetic Data with Law Enforcement}
When controlling for demographic changes, we see that acceptability of genetic data sharing with law enforcement did not change significantly from 2019 to 2020 or 2020 to 2021. However the overall effect was a steady decline from 2019 to 2021 such that the difference between these two years shows a significant decrease.

Before the onset of the pandemic, we observe the following demographic effects: men are significantly less likely to find sharing genetic data with law enforcement acceptable; age positively correlates with acceptance --- O.R. for 50-64 is 1.77 and O.R. for 65+ is 2.16; and education negatively correlates with acceptance --- O.R. for some college is  0.78  and O.R. for a Bachelor's degree or more is 0.73. 

After the pandemic's onset in 2020, we see significant demographic shifts in responses, again leading to fewer within-group differences. Specifically, older individuals, who were generally more likely to agree with the statement became just as likely as younger individuals to find it acceptable to share genetic data with law enforcement. Further, men, who in 2019 were less likely than women to think genetic data sharing with law enforcement was acceptable, became just as likely as women in 2020. The two demographic shifts that occurred in 2020, age and gender, both held firm into 2021 when compared to their 2019 levels.

Finally, we again observe a change between respondents with differing political affiliations. In 2020, all demographics do not show disparities between their different levels, except for political party -- Republican-leaning individuals were more likely to find sharing genetic data with law enforcement acceptable. However, in 2021, this trend reversed and Republican-leaning individuals were no longer more likely than Democratic-leaning individuals to find it acceptable to share genetic data with law enforcement.

\subsection{Fitness Data Sharing with Medical Researchers}
Overall, we see that the acceptability of sharing fitness data with medical researchers increased from  2019 into 2020 and remained at a higher level in 2021.

In 2019, age negatively correlated with acceptance of data sharing with medical researchers. Further, Hispanic individuals were slightly more likely to find this practice acceptable. No other demographic groups exhibited internal disparities in acceptance for this item in 2019.  

However, after the onset of the pandemic, we saw significant changes in acceptability between different demographic groups, again leading to fewer overall within-group differences. First, we see that older individuals became as accepting as younger individuals. An increase also occurred for those who have a Bachelor's degree. Acceptability for those who identify as Hispanic decreased to the level of non-Hispanics.

\begin{table*}[!htbp] \centering 
  \caption{Inter-year logistic regression models for the relationship between acceptance of a particular data use scenario and sociodemographics. Odds ratios are shown and * depict significance of relationship. Interaction between years shown; fixed effects are omitted for brevity. For example, the first column depicts the comparison between 2019 and 2020 for the Government/Terrorism question, and a significant value reported indicates that there was a significant change for that sociodemographic variable in 2020 when compared to 2019 --- in this case, men and those with a Bachelor's degree or more were significantly more likely to report acceptance of this data use in 2020 compared to 2019.} 
  \label{tbl:inter_years} 
\resizebox{\linewidth}{!}{
\begin{tabular}{@{\extracolsep{5pt}}lcccccccccccc} 
\\[-1.8ex]\hline 
\hline \\[-1.8ex] 
 & \multicolumn{3}{c}{\textit{Item 1:}} & \multicolumn{3}{c}{\textit{Item 2:}} & \multicolumn{3}{c}{\textit{Item 3:}} & \multicolumn{3}{c}{\textit{Item 4:}} \\ 
 & \multicolumn{3}{c}{\textit{Government/Terrorism}} & \multicolumn{3}{c}{\textit{Law Enforcement/Genetic}} & \multicolumn{3}{c}{\textit{Medical Research/Fitness}} & \multicolumn{3}{c}{\textit{Corporate/Mental Health}}  \\ 
\cline{2-4} \cline{5-7} \cline{8-10} \cline{11-13} \\
 & 2019-20 & 2020-21 & 2019-21 & 2019-20 & 2020-21 & 2019-21 & 2019-20 & 2020-21 & 2019-21 & 2019-20 & 2020-21 & 2019-21 \\ 
\hline \\[-1.8ex] 
  Gender:Man * YEAR  &  1.362$^{*}$  & 0.974 &  1.326$^{*}$  &  1.717$^{**}$  & 0.828 &  1.422$^{*}$  & 1.265 & 1.121 &  1.418$^{*}$  &  1.563$^{**}$  & 1.032 &  1.614$^{**}$ \\ 
   &   &   &   &   &   &   &   &   &   &   &   &  \\ 
 Age: 30-49 * YEAR  & 1.203 & 0.979 & 1.178 & 0.813 & 1.16 & 0.944 & 1.468 & 0.812 & 1.191 &  2.119$^{**}$  & 1.006 &  2.131$^{**}$ \\ 
   &   &   &   &   &   &   &   &   &   &   &   &  \\ 
 Age: 50-64 * YEAR  & 0.785 & 1.126 & 0.884 &  0.408$^{**}$  & 1.592 &  0.650$^{*}$  &  1.829$^{*}$ & 1.401 &  2.562$^{**}$  &  1.650$^{*}$  & 0.98 &  1.616$^{*}$ \\ 
   &   &   &   &   &   &   &   &   &   &   &   &  \\ 
 Age: 65+ * YEAR  & 0.695 & 0.975 & 0.677 &  0.406$^{**}$  & 1.526 &  0.620$^{*}$  &  1.890$^{*}$  & 1.524 &  2.880$^{**}$  &  2.141$^{**}$  & 0.883 &  1.891$^{**}$ \\ 
   &   &   &   &   &   &   &   &   &   &   &   &  \\ 
 Race: Black * YEAR  & 0.835 & 1.397 & 1.166 & 1.101 & 1.238 & 1.363 & 0.783 & 0.944 & 0.739 & 0.781 & 0.882 &  0.689 \\ 
   &   &   &   &   &   &   &   &   &   &   &   &  \\ 
 Race: Asian * YEAR  & 0.729 & 0.898 & 0.655 & 1.059 & 1.072 & 1.135 & 0.672 & 1.49 & 1.001 & 0.871 & 0.921 &  0.802 \\ 
   &   &   &   &   &   &   &   &   &   &   &   &  \\ 
 Race: Other * YEAR  & 1.524 & 0.808 & 1.232 & 1.14 & 1.144 & 1.304 & 1.07 & 0.907 & 0.97 & 1.074 & 1.047 &  1.124 \\ 
   &   &   &   &   &   &   &   &   &   &   &   &  \\ 
 Ethn.: Hispanic * YEAR  & 0.837 &  1.844$^{*}$  & 1.543 & 0.82 & 1.46 & 1.197 &  0.524$^{*}$  & 1.443 & 0.756 & 0.914 & 1.425 &  1.303 \\ 
&   &   &   &   &   &   &   &   &   &   &   &  \\ 
 Ed. Att.: Some college * YEAR  & 1.412 & 0.825 & 1.165 & 1.005 & 1.207 & 1.214 & 1.329 & 0.958 & 1.273 & 1.475 & 0.773 &  1.139 \\ 
   &   &   &   &   &   &   &   &   &   &   &   &  \\ 
 Ed. Att.: Bachelor's+ * YEAR  &  1.586$^{*}$  & 0.888 & 1.409 & 1.259 & 1.071 & 1.348 &  1.824$^{**}$  & 0.852 &  1.554$^{*}$  &  1.706$^{**}$  & 0.777 &  1.327 \\ 
   &   &   &   &   &   &   &   &   &   &   &   &  \\ 
 Political Id.: Rep/Lean Rep * YEAR  & 0.94 &  0.522$^{**}$  &  0.490$^{**}$  & 1.129 &  0.642$^{*}$  &  0.725$^{*}$  & 1.239 &  0.497$^{**}$  &  0.615$^{**}$  & 1.176 & 0.886 &  1.041 \\ 
   &   &   &   &   &   &   &   &   &   &   &   &  \\ 
 Intercept  & 1.014 &  0.571$^{**}$  & 1.014 & 1.006 & 1.006 & 1.006 & 1.191 & 0.687 & 1.191 & 1.099 & 0.802 &  1.099 \\ 
   &   &   &   &   &   &   &   &   &   &   &   &  \\ 
\hline 
\hline \\[-1.8ex] 
\textit{Note:}  &&&&&&&&&\multicolumn{4}{r}{$^{*}$p $<$ 0.05; $^{**}$p $<$ 0.01} \\ 
\end{tabular} }
\end{table*} 

The demographic changes brought on by the pandemic lasted into 2021 for both age and education. Being over 65, which in 2019 {\it negatively} correlated with acceptability, in 2021 was {\it positively} correlated with acceptability of sharing data with medical researchers. We hypothesize this change is due to sustained higher COVID-19 risk and concern -- even with vaccines that came out in 2021 -- among those 65+. Concerns about the pandemic may have generalized to general increases in acceptability of medical research for other conditions at which they are high risk (i.e., heart disease) among older respondents~\cite{centers2020older,HeartHea90:online}. The change in sentiment among Bachelor's degree holders also holds in 2021: there was no education effect in 2019, but those a Bachelor's degree viewed this data use as more acceptable in 2021 as well as 2020 than their less-educated peers, who may have become more informed about the role of medical research in fighting various conditions as a result of the pandemic~\cite{rattay2021differences}. 

Gender, which saw no difference in acceptability in 2019, saw increases for men in 2020, and in 2021 that increase rose to the level of statistical significance from 2019. 

Like the previous two items, we also saw a marked change in the political party disparity in 2021. Whereas there was no significant difference in acceptability views for 2019 and 2020, Republican-leaning identified individuals became significantly less likely to accept sharing fitness data than Democratic-leaning individuals in 2021 both when compared to 2019 and 2020.

\subsection{Monitoring for Mental Health Crises by Social Media Companies}
Finally, we see that, overall, the acceptability of a social media company monitoring posts for mental health crises increased significantly and with large magnitude between 2019 and 2020 and remained at a higher level in 2021.

In 2019, we found all individuals over 30 were less accepting of the idea of social media companies monitoring data for mental health crises compared to the youngest group. Additionally, we see that Bachelor's degree holders and Republican-leaning individuals are less likely to find this behavior acceptable in 2019.

After the onset of the pandemic, we see meaningful changes amongst demographic groups. First, those in the 30-49 group became as likely as the youngest group to find this acceptable, a statistically significant change. Older individuals in the 50-64 and 65+ groups also saw a relative increase in their thoughts on acceptability, though older people still found social media monitoring for mental health crises to be less acceptable than younger people; this difference between years was significant. Further, men, who in 2019 had similar levels of acceptance of this data use to women, became more likely than women in 2020 to find this practice acceptable, another significant change in sentiment. This is also the case for those with a Bachelor's degree, who in 2019 were less likely to find social media companies monitoring for mental health crises acceptable but in 2020 were more likely to find it acceptable.

The demographic changes for age and gender both held into 2021. When compared to 2019, older individuals became more accepting of this practice. Men also continued to find social media monitoring for mental health crises was acceptable at higher rates than women in 2021, a significant change from 2019. However, the significant change from 2019 to 2020 we saw for those with a Bachelor's degree receded in 2021; in 2021, like 2019, a person with a Bachelor's degree was no more likely than those with less education to find social media monitoring for mental health crises acceptable.
\section{Discussion}
This work seeks to understand changes in U.S. public sentiment toward government- and health-related data use after the onset (RQ1) and continuation (RQ2) of the COVID-19 pandemic, as well as how sociodemographics relate to these changes (RQ3). To answer these questions, we measure acceptance of four concrete privacy-related data use scenarios in 2019, 2020, and 2021. 

In line with findings from prior work \cite{westin2003social,best2006privacy,biddle2022data} that data privacy sentiments in contexts closely related to major events shift in tandem with those events, we observe significant shifts in acceptability in all four data use scenarios we investigate. Answering RQ1, between the years 2019 and 2020, we observe a \textit{decrease} in acceptability in government collection of data on Americans to assess terrorism threats and an \textit{increase} in acceptability of both health scenarios: sharing user data from a fitness tracking app with medical researchers studying the link between exercise and heart disease; and a social media company monitoring its users’ posts for signs of depression to identify people who are at risk of self-harm and connect them to counseling services.

Answering RQ2, we find that these changes sustained from 2020 to 2021 and that by 2021, there was also a \textit{decrease} in acceptability of the other government-related data use scenario: DNA testing companies sharing their customers’ genetic data with law enforcement agencies in order to help solve crimes. 

Finally, in answer to RQ3, we find differences in these shifts depending on the sociodemographic identity of the respondent: respondents' political identity, and the alignment between that identity and the political party in power, played an increasingly significant role in the privacy attitudes we measured.

Here, we discuss hypotheses for the causes of these shifts, which serve as directions for future work seeking to understand why people feel the way they do about data privacy and how their feelings change. We acknowledge that our analysis is correlational in nature rather than causal; similarly, we acknowledge that factors such as our sampling methods or individual differences may have contributed to patterns we observe in our data. Thus, such future work is necessary to confirm or refute the causes we hypothesize.

We hypothesize that acceptance of government data uses may relate to decreased trust in both the government and law enforcement and law enforcement data uses may have decreased due to  as well as decreased trust in law enforcement between 2019 and 2021~\cite{howard,Pew01,nytblm,cohn2020public}. 
Prior work finds that trust is related to how much an individual views the government or security technologies as threatening~\cite{anthony2015big,pavone2012public}. Further, prior examinations of privacy sentiment over time \cite{westin2003social,best2006privacy} found that erosion of trust due to events such as the Watergate scandal, Snowden leaks, and corporate violations with the growth of the internet in the mid-1990s, all led to increases in concerns about digital surveillance. Similar to these prior events, trust in the U.S. federal government declined significantly over the course of the pandemic~\cite{pewgov} as did reduced perceptions of the U.S. government as organized, clear, and knowledgeable in response to COVID-19~\cite{kim2020analysis,han2021trust}. As a result, respondents may have lowered their trust in and consequently acceptance of government use of personal data for public safety purposes such as preventing terrorism. 

Similarly, trust in law enforcement decreased by 2$\times$ any previous yearly decreases by June 2020~\cite{cohn2020public} after the rise in salience of the Black Lives Matter movement, which first started in 2013 in response to police brutality and other systemic issues impacting Black individuals but gained widespread visibility with international protests occurring in the summer of 2020~\cite{howard,Pew01}. In line with prior work finding that this disapproval has led to reduced citizen engagement in public safety and reporting of crimes~\cite{ang2021police}, we hypothesize that by 2021 reduced trust in law enforcement led many respondents to no longer accept the sharing of genetic data with law enforcement.

Together, these findings support both CI theory and prior empirical work (including in the context of COVID-19) suggesting that trust in the entity providing a technology~\cite{hargittai2020americans} or using data~\cite{majumder2016beyond} strongly influences people's willingness to share their data or adopt technologies. Importantly, we highlight that such trust is dynamic: changes in entity trust may lead users to stop using a technology they previously adopted, or no longer be willing to share data they once happily contributed. Our results suggest that such changes, which can have significant impact on technology adoption and thus design, may be predictable by tracking trust metrics.

We note that while these hypotheses focus on the data stakeholders, we acknowledge that participants may have judged our chosen scenarios based on the data uses or data types instead of, or in addition to, the data stakeholders. For example, when assessing the scenario regarding government data use, it is possible that respondents put more focus on the purpose of data collection -- in this case, assessing terrorist threats -- than on the government as a stakeholder in the data collection process. Once the spread of infectious diseases grew to be the top threat for people in the U.S. in 2020 and surpassed the concern for terrorism~\cite{pewdisease}, this re-prioritization of threats may have led fewer people in the U.S. to find this data use acceptable, thus providing an alternative explanation for our results. We encourage future work to isolate participants' attitudes towards different data types vs. data uses vs. stakeholders; doing so would further our understanding of triggers associated with changes in data privacy sentiment.

Regarding respondents' increased approval toward fitness tracker data use for medical research, we hypothesize that the focus on COVID-19 increased the apparent relevance of medical research, even on other diseases. Indeed, the "spread of infectious diseases" became the top perceived national threat by people in 2020~\cite{pewdisease}; it is possible that the medical and social benefits of sharing data became more important than privacy to respondents during this time. This concern regarding medical issues may have also increased people's acceptance of the use of their social media data for detecting and intervening in mental health issues. Further, contextual factors such as the increase in mental health issues~\cite{pfefferbaum2020mental} and social media use~\cite{kemp2020digital}, respectively, as well as mental health issues caused by social media use~\cite{zhao2020social} during the pandemic, may have further increased people in the U.S.'s willingness to allow their social media data to be used for mental health purposes.

These shifts suggest that people may be more willing to adopt health technologies and contribute their digital data for a variety of health purposes than ever before. While this offers exciting directions for the development of new health technologies, future work and design must also carefully consider how to ensure informed consent and appropriate data transparency for these technologies that people may be inclined to quickly adopt without consideration of potential consequences.

On the whole, many of the strong differences in data privacy sentiment within demographic groups that we observed in 2019 are not present in 2020 or 2021. This point is especially true for government and law enforcement data uses: When controlling for other variables, we note that women, older people, and less educated people were all more likely to find these data uses acceptable compared to their counterparts in 2019. Confirming the direction of the sentiment changes with the raw data, we see that these groups decrease their acceptance of government and law enforcement data uses, so that in 2020 differences between in-group members disappeared. 

In the wake of an increasingly politically polarized landscape in the U.S.~\cite{pew2019politics}, there is a growing body of literature investigating political partisanship's relationship with attitudes, behaviors and self-identification~\cite{huddy2017political,greene1999understanding,west2020partisanship}. Party membership is one of the most important predictor variables for political attitudes and behaviors~\cite{gerber2010party}, and, increasingly, attitudes and behaviors that are not directly connected to political processes. Partisanship is among the strongest predictors of attitudes toward topics ranging from public health-related attitudes and behaviors during the early stages of the COVID-19 pandemic~\cite{gadarian2021partisanship,grossman2020political} to perceptions of fairness in algorithmic decision-making~\cite{grgic2020dimensions}. People's behaviors are also known cycle based on whether their political party is in power~\cite{chapman2022rage}; gun sales increase during Democratic presidential terms~\cite{depetris2015fear} and donations to women's health and progressive law organizations increase during Republican presidential terms \cite{insidephilanthropy2021,atlantic2016}. 

In line with these findings, we find significant differences in Republican and Democrat data privacy sentiments throughout our data collection period. In 2019 and 2020, we find that Republicans had significantly higher acceptance of government data collection for terrorism assessment compared to Democrats; however, by 2021, their acceptance dropped to be significantly lower than Democrats. We also observe Republicans becoming less accepting of sharing their genetic data with law enforcement between 2019 and 2021.

There are few surveys that explore partisan differences in data privacy attitudes in detail; sentiment is usually assessed by capturing Democrat and Republican views on privacy-related topics at singular points in time (e.g.,~\cite{auxier2019americans,reddick2015public,pew2016more}). Recent work finds conservative Republicans to be associated with warmer attitudes towards surveillance, and liberalism to be associated with less acceptance of government surveillance compared to conservatism~\cite{turow2018divided, nam2019determines}. Our work alludes to this possibly being a condition of the current Presidential administration, rather than an inherent trait of either political affiliation.

While our work cannot draw causal conclusions, we note the presidential election -- in which Republican candidate Donald Trump finished his term and was succeeded by Democratic candidate Joe Biden~\cite{nytpres} -- that occurred before our third round of data collection as a possible reason for the stark decrease in acceptance in government and law enforcement data uses among Republicans. We hypothesize that -- like gun sales and donations to particular causes -- views on data uses by governments may relate not to sentiments toward ``the government'' at large, but rather to fear of what the opposing party~\cite{chapman2022rage} may do with their data. If so, changes in privacy sentiment -- which significantly influence legislation such as that around government data uses like end-to-end encryption~\cite{citron2016privacy,pew2016more} -- may be easily predictable. We encourage future work to confirm the relationship between an individual's sentiment toward government data uses and the alignment between their political identity and the political party in power. 
\section{Conclusion}
We examined the data privacy sentiments of those in the U.S. between 2019 and 2021 using repeated cross-sectional surveys that measured their acceptance of four concrete data use scenarios. We find that following the onset of the COVID-19 pandemic, respondent acceptance of government collection of data on Americans to assess terrorism threats decreased, while their acceptance of health-related data use increased for both 1) use of fitness tracker data by medical researchers studying the link between exercise and heart disease, and 2) use of social media data by a social media company to detect and intervene in users' mental health issues. In 2021, we observe that the 2020 changes in sentiment are sustained, and that respondent acceptance of law enforcement use of genetic data for crime detection decreased when compared to 2019. Together, these findings suggest that data privacy sentiments may in fact change in tandem with major geopolitical and national events, and the effects may have broad impacts to multiple privacy contexts. 

While we find that sentiments became more cohesive across demographic groups during the pandemic, one notable exception to this finding is sentiment within political affiliation groups, which appear to change in tandem with the changing of the political party in power, for example, after the 2020 national election.

At the time of this writing the COVID-19 pandemic continues to progress. We encourage future research on data privacy sentiments as they may continue to change throughout and after the pandemic, offering insight into the changing landscape of public opinion into which new privacy-sensitive technologies and policies may be introduced.

\section*{Acknowledgements}
A portion of this work was done while the third author was at Microsoft Research. The authors also wish to thank Microsoft Research for supporting the 2020 wave of data collection.

{\footnotesize \bibliographystyle{acm}
\bibliography{references}}

\clearpage
\appendix

\begin{table*}[!htbp] \centering 
  \caption{Question 1 (government data collection for terrorism threat assessment) regressions for each year separately 2019, 2020, and 2021} 
  \label{tbl:QuestionA_years} 
  \resizebox{!}{9cm}{
\begin{tabular}{@{\extracolsep{5pt}}lccc} 
\\[-1.8ex]\hline 
\hline \\[-1.8ex]

 & 2019 & 2020 & 2021 \\ 
\\[-1.8ex] & (1) & (2) & (3)\\ 
\hline \\[-1.8ex] 
 GENDERMale & 0.717 & 0.977 & 0.951 \\ 
  & (0.598, 0.861) & (0.762, 1.251) & (0.769, 1.177) \\ 
  & p = 0.0004$^{**}$ & p = 0.853 & p = 0.645 \\ 
  & & & \\ 
 AGECAT30-49 & 1.045 & 1.257 & 1.231 \\ 
  & (0.793, 1.378) & (0.904, 1.749) & (0.941, 1.611) \\ 
  & p = 0.755 & p = 0.174 & p = 0.131 \\ 
  & & & \\ 
 AGECAT50-64 & 1.353 & 1.062 & 1.196 \\ 
  & (1.023, 1.790) & (0.733, 1.538) & (0.861, 1.661) \\ 
  & p = 0.035$^{*}$ & p = 0.752 & p = 0.287 \\ 
  & & & \\ 
 AGECAT65+ & 1.437 & 0.998 & 0.973 \\ 
  & (1.065, 1.938) & (0.660, 1.508) & (0.688, 1.376) \\ 
  & p = 0.018$^{*}$ & p = 0.993 & p = 0.877 \\ 
  & & & \\ 
 RACEBlack or African American & 0.986 & 0.823 & 1.150 \\ 
  & (0.730, 1.331) & (0.592, 1.144) & (0.855, 1.546) \\ 
  & p = 0.926 & p = 0.246 & p = 0.356 \\ 
  & & & \\ 
 RACEAsian or Asian-American & 1.640 & 1.195 & 1.073 \\ 
  & (0.933, 2.883) & (0.805, 1.775) & (0.747, 1.542) \\ 
  & p = 0.086 & p = 0.378 & p = 0.702 \\ 
  & & & \\ 
 RACEOther & 0.667 & 1.017 & 0.822 \\ 
  & (0.465, 0.958) & (0.678, 1.527) & (0.522, 1.297) \\ 
  & p = 0.029$^{*}$ & p = 0.934 & p = 0.401 \\ 
  & & & \\ 
 HISPANICYes & 1.124 & 0.941 & 1.735 \\ 
  & (0.845, 1.496) & (0.633, 1.398) & (1.118, 2.691) \\ 
  & p = 0.422 & p = 0.763 & p = 0.014$^{*}$ \\ 
  & & & \\ 
 EDUCATIONBachelor's or more & 0.735 & 1.166 & 1.036 \\ 
  & (0.592, 0.913) & (0.869, 1.566) & (0.793, 1.353) \\ 
  & p = 0.006$^{**}$ & p = 0.307 & p = 0.795 \\ 
  & & & \\ 
 EDUCATIONSome college & 0.801 & 1.131 & 0.933 \\ 
  & (0.636, 1.007) & (0.830, 1.541) & (0.711, 1.223) \\ 
  & p = 0.058 & p = 0.437 & p = 0.615 \\ 
  & & & \\ 
 POLPARTYRep/Lean Rep & 1.518 & 1.427 & 0.745 \\ 
  & (1.253, 1.841) & (1.101, 1.848) & (0.595, 0.932) \\ 
  & p = 0.00003$^{**}$ & p = 0.008$^{**}$ & p = 0.011$^{*}$ \\ 
  & & & \\ 
 Constant & 1.014 & 0.571 & 0.658 \\ 
  & (0.755, 1.363) & (0.380, 0.857) & (0.484, 0.895) \\ 
  & p = 0.925 & p = 0.007$^{**}$ & p = 0.008$^{**}$ \\ 
  & & & \\ 
\hline \\[-1.8ex] 
Observations & 2,012 & 1,138 & 1,537 \\ 
Log Likelihood & $-$1,361.237 & $-$769.597 & $-$1,026.828 \\ 
Akaike Inf. Crit. & 2,746.474 & 1,563.193 & 2,077.655 \\ 
\hline 
\hline \\[-1.8ex] 
\textit{Note:}  & \multicolumn{3}{r}{$^{*}$p$<$0.05; $^{**}$p$<$0.01} \\ 
\end{tabular} }
\end{table*} 

\begin{table*}[!htbp] \centering 
  \caption{Question 2 (sharing genetic data with law enforcement) regressions for each year separately 2019, 2020, and 2021} 
  \label{tbl:QuestionC_years} 
  \resizebox{!}{11cm}{
\begin{tabular}{@{\extracolsep{5pt}}lccc} 
\\[-1.8ex]\hline 
\hline \\[-1.8ex]

 & 2019 & 2020 & 2021 \\ 
\\[-1.8ex] & (1) & (2) & (3)\\ 
\hline \\[-1.8ex] 
 GENDERMale & 0.633 & 1.087 & 0.900 \\ 
  & (0.527, 0.761) & (0.850, 1.390) & (0.730, 1.111) \\ 
  & p = 0.00001$^{**}$ & p = 0.505 & p = 0.327 \\ 
  & & & \\ 
 AGECAT30-49 & 1.191 & 0.969 & 1.124 \\ 
  & (0.902, 1.574) & (0.700, 1.342) & (0.862, 1.467) \\ 
  & p = 0.218 & p = 0.849 & p = 0.389 \\ 
  & & & \\ 
 AGECAT50-64 & 1.765 & 0.721 & 1.147 \\ 
  & (1.332, 2.340) & (0.499, 1.041) & (0.829, 1.587) \\ 
  & p = 0.0001$^{**}$ & p = 0.081 & p = 0.407 \\ 
  & & & \\ 
 AGECAT65+ & 2.157 & 0.876 & 1.337 \\ 
  & (1.594, 2.919) & (0.584, 1.315) & (0.954, 1.874) \\ 
  & p = 0.00000$^{**}$ & p = 0.524 & p = 0.092 \\ 
  & & & \\ 
 RACEBlack or African American & 0.843 & 0.928 & 1.149 \\ 
  & (0.623, 1.142) & (0.672, 1.283) & (0.856, 1.542) \\ 
  & p = 0.270 & p = 0.653 & p = 0.355 \\ 
  & & & \\ 
 RACEAsian or Asian-American & 0.847 & 0.897 & 0.962 \\ 
  & (0.480, 1.495) & (0.604, 1.333) & (0.671, 1.380) \\ 
  & p = 0.568 & p = 0.593 & p = 0.834 \\ 
  & & & \\ 
 RACEOther & 0.805 & 0.917 & 1.049 \\ 
  & (0.562, 1.152) & (0.612, 1.374) & (0.673, 1.636) \\ 
  & p = 0.236 & p = 0.675 & p = 0.834 \\ 
  & & & \\ 
 HISPANICYes & 1.012 & 0.829 & 1.211 \\ 
  & (0.760, 1.347) & (0.559, 1.230) & (0.785, 1.867) \\ 
  & p = 0.938 & p = 0.353 & p = 0.387 \\ 
  & & & \\ 
 EDUCATIONBachelor's or more & 0.730 & 0.919 & 0.984 \\ 
  & (0.587, 0.908) & (0.687, 1.230) & (0.756, 1.281) \\ 
  & p = 0.005$^{**}$ & p = 0.571 & p = 0.906 \\ 
  & & & \\ 
 EDUCATIONSome college & 0.784 & 0.788 & 0.951 \\ 
  & (0.623, 0.987) & (0.580, 1.071) & (0.729, 1.242) \\ 
  & p = 0.039$^{*}$ & p = 0.128 & p = 0.714 \\ 
  & & & \\ 
 POLPARTYRep/Lean Rep & 1.164 & 1.314 & 0.844 \\ 
  & (0.959, 1.412) & (1.015, 1.700) & (0.677, 1.052) \\ 
  & p = 0.124 & p = 0.039$^{*}$ & p = 0.131 \\ 
  & & & \\ 
 Constant & 1.006 & 1.006 & 0.754 \\ 
  & (0.748, 1.355) & (0.674, 1.500) & (0.557, 1.022) \\ 
  & p = 0.967 & p = 0.978 & p = 0.069 \\ 
  & & & \\ 
\hline \\[-1.8ex] 
Observations & 2,012 & 1,138 & 1,537 \\ 
Log Likelihood & $-$1,352.573 & $-$778.105 & $-$1,049.512 \\ 
Akaike Inf. Crit. & 2,729.146 & 1,580.209 & 2,123.024 \\ 
\hline 
\hline \\[-1.8ex] 
\textit{Note:}  & \multicolumn{3}{r}{$^{*}$p$<$0.05; $^{**}$p$<$0.01} \\ 
\end{tabular} }
\end{table*}

\begin{table*}[!htbp] \centering 
  \caption{Question 3 (fitness data sharing with medical researchers) regressions for each year separately 2019, 2020, and 2021} 
  \label{tbl:QuestionD_years} 
  \resizebox{!}{11cm}{
\begin{tabular}{@{\extracolsep{5pt}}lccc} 
\\[-1.8ex]\hline 
\hline \\[-1.8ex]

 & 2019 & 2020 & 2021 \\ 
\\[-1.8ex] & (1) & (2) & (3)\\ 
\hline \\[-1.8ex] 
 GENDERMale & 0.971 & 1.228 & 1.377 \\ 
  & (0.806, 1.170) & (0.960, 1.572) & (1.114, 1.704) \\ 
  & p = 0.758 & p = 0.102 & p = 0.004$^{**}$ \\ 
  & & & \\ 
 AGECAT30-49 & 0.697 & 1.023 & 0.831 \\ 
  & (0.528, 0.922) & (0.738, 1.420) & (0.635, 1.088) \\ 
  & p = 0.012$^{*}$ & p = 0.890 & p = 0.178 \\ 
  & & & \\ 
 AGECAT50-64 & 0.442 & 0.809 & 1.134 \\ 
  & (0.332, 0.590) & (0.560, 1.170) & (0.817, 1.573) \\ 
  & p = 0.00000$^{**}$ & p = 0.261 & p = 0.454 \\ 
  & & & \\ 
 AGECAT65+ & 0.513 & 0.969 & 1.477 \\ 
  & (0.378, 0.696) & (0.645, 1.457) & (1.048, 2.082) \\ 
  & p = 0.00002$^{**}$ & p = 0.881 & p = 0.026$^{*}$ \\ 
  & & & \\ 
 RACEBlack or African American & 1.127 & 0.882 & 0.833 \\ 
  & (0.824, 1.541) & (0.638, 1.220) & (0.617, 1.122) \\ 
  & p = 0.454 & p = 0.449 & p = 0.230 \\ 
  & & & \\ 
 RACEAsian or Asian-American & 1.461 & 0.982 & 1.463 \\ 
  & (0.881, 2.424) & (0.661, 1.460) & (1.013, 2.113) \\ 
  & p = 0.143 & p = 0.930 & p = 0.043$^{*}$ \\ 
  & & & \\ 
 RACEOther & 0.912 & 0.975 & 0.884 \\ 
  & (0.631, 1.317) & (0.649, 1.465) & (0.564, 1.386) \\ 
  & p = 0.623 & p = 0.904 & p = 0.593 \\ 
  & & & \\ 
 HISPANICYes & 1.361 & 0.713 & 1.029 \\ 
  & (1.007, 1.840) & (0.479, 1.061) & (0.664, 1.594) \\ 
  & p = 0.046$^{*}$ & p = 0.096 & p = 0.899 \\ 
  & & & \\ 
 EDUCATIONBachelor's or more & 0.952 & 1.738 & 1.481 \\ 
  & (0.764, 1.188) & (1.295, 2.331) & (1.132, 1.936) \\ 
  & p = 0.666 & p = 0.0003$^{**}$ & p = 0.005$^{**}$ \\ 
  & & & \\ 
 EDUCATIONSome college & 1.023 & 1.360 & 1.302 \\ 
  & (0.810, 1.292) & (0.999, 1.850) & (0.993, 1.708) \\ 
  & p = 0.848 & p = 0.051 & p = 0.057 \\ 
  & & & \\ 
 POLPARTYRep/Lean Rep & 0.848 & 1.050 & 0.522 \\ 
  & (0.696, 1.033) & (0.810, 1.361) & (0.417, 0.653) \\ 
  & p = 0.102 & p = 0.713 & p = 0.000$^{**}$ \\ 
  & & & \\ 
 Constant & 1.191 & 0.687 & 0.778 \\ 
  & (0.881, 1.610) & (0.459, 1.028) & (0.572, 1.058) \\ 
  & p = 0.257 & p = 0.068 & p = 0.110 \\ 
  & & & \\ 
\hline \\[-1.8ex] 
Observations & 1,989 & 1,138 & 1,537 \\ 
Log Likelihood & $-$1,314.534 & $-$774.020 & $-$1,026.291 \\ 
Akaike Inf. Crit. & 2,653.068 & 1,572.040 & 2,076.582 \\ 
\hline 
\hline \\[-1.8ex] 
\textit{Note:}  & \multicolumn{3}{r}{$^{*}$p$<$0.05; $^{**}$p$<$0.01} \\ 
\end{tabular} }
\end{table*}

\begin{table*}[!htbp] \centering 
  \caption{Question 4 (monitoring for mental health crises by social media companies) regressions for each year separately 2019, 2020, and 2021} 
  \label{tbl:QuestionE_years} 
  \resizebox{!}{11cm}{
\begin{tabular}{@{\extracolsep{5pt}}lccc} 
\\[-1.8ex]\hline 
\hline \\[-1.8ex]

 & 2019 & 2020 & 2021 \\ 
\\[-1.8ex] & (1) & (2) & (3)\\ 
\hline \\[-1.8ex] 
 GENDERMale & 0.848 & 1.326 & 1.369 \\ 
  & (0.686, 1.049) & (1.034, 1.700) & (1.106, 1.694) \\ 
  & p = 0.129 & p = 0.027$^{**}$ & p = 0.004$^{***}$ \\ 
  & & & \\ 
 AGECAT30-49 & 0.473 & 1.002 & 1.007 \\ 
  & (0.353, 0.632) & (0.723, 1.387) & (0.773, 1.312) \\ 
  & p = 0.00000$^{***}$ & p = 0.993 & p = 0.958 \\ 
  & & & \\ 
 AGECAT50-64 & 0.297 & 0.490 & 0.480 \\ 
  & (0.218, 0.405) & (0.337, 0.713) & (0.344, 0.672) \\ 
  & p = 0.000$^{***}$ & p = 0.0002$^{***}$ & p = 0.00002$^{***}$ \\ 
  & & & \\ 
 AGECAT65+ & 0.304 & 0.652 & 0.576 \\ 
  & (0.218, 0.426) & (0.433, 0.982) & (0.409, 0.810) \\ 
  & p = 0.000$^{***}$ & p = 0.041$^{**}$ & p = 0.002$^{***}$ \\ 
  & & & \\ 
 RACEBlack or African American & 1.264 & 0.987 & 0.871 \\ 
  & (0.903, 1.770) & (0.712, 1.368) & (0.646, 1.173) \\ 
  & p = 0.173 & p = 0.938 & p = 0.364 \\ 
  & & & \\ 
 RACEAsian or Asian-American & 1.280 & 1.114 & 1.026 \\ 
  & (0.741, 2.211) & (0.748, 1.660) & (0.716, 1.470) \\ 
  & p = 0.377 & p = 0.595 & p = 0.890 \\ 
  & & & \\ 
 RACEOther & 0.774 & 0.831 & 0.870 \\ 
  & (0.511, 1.173) & (0.551, 1.255) & (0.555, 1.365) \\ 
  & p = 0.228 & p = 0.380 & p = 0.546 \\ 
  & & & \\ 
 HISPANICYes & 1.085 & 0.992 & 1.413 \\ 
  & (0.776, 1.517) & (0.665, 1.478) & (0.912, 2.191) \\ 
  & p = 0.634 & p = 0.967 & p = 0.123 \\ 
  & & & \\ 
 EDUCATIONBachelor's or more & 0.772 & 1.317 & 1.023 \\ 
  & (0.601, 0.991) & (0.979, 1.771) & (0.784, 1.337) \\ 
  & p = 0.043$^{**}$ & p = 0.069$^{*}$ & p = 0.865 \\ 
  & & & \\ 
 EDUCATIONSome college & 0.884 & 1.303 & 1.006 \\ 
  & (0.680, 1.147) & (0.955, 1.778) & (0.768, 1.319) \\ 
  & p = 0.354 & p = 0.096$^{*}$ & p = 0.963 \\ 
  & & & \\ 
 POLPARTYRep/Lean Rep & 0.748 & 0.880 & 0.779 \\ 
  & (0.596, 0.938) & (0.677, 1.143) & (0.624, 0.973) \\ 
  & p = 0.013$^{**}$ & p = 0.338 & p = 0.028$^{**}$ \\ 
  & & & \\ 
 Constant & 1.099 & 0.802 & 0.942 \\ 
  & (0.798, 1.513) & (0.536, 1.200) & (0.695, 1.277) \\ 
  & p = 0.562 & p = 0.284 & p = 0.701 \\ 
  & & & \\ 
\hline \\[-1.8ex] 
Observations & 1,989 & 1,138 & 1,537 \\ 
Log Likelihood & $-$1,089.413 & $-$765.412 & $-$1,030.537 \\ 
Akaike Inf. Crit. & 2,202.825 & 1,554.824 & 2,085.073 \\ 
\hline 
\hline \\[-1.8ex] 
\textit{Note:}  & \multicolumn{3}{r}{$^{*}$p$<$0.05; $^{**}$p$<$0.01} \\ 
\end{tabular} }
\end{table*}

\begin{table*}[!htbp] \centering 
  \caption{Interaction between 2019 and 2020, with 2019 as reference. Fixed effects are omitted for brevity. See Table~\ref{tbl:QuestionA_years}-\ref{tbl:QuestionE_years} for reference.} 
  \label{tbl:model_2019_2020} 
  \resizebox{\linewidth}{11cm}{
\begin{tabular}{@{\extracolsep{5pt}}lcccc} 
\\[-1.8ex]\hline 
\hline \\[-1.8ex]

 & Question 1 & Question 2 & Question 3 & Question 4 \\ 
 & Government & Law Enforcement & Medical Research & Corporate \\ 
 & Terrorism & Genetic & Fitness & Mental Health \\ 
\\[-1.8ex] & (1) & (2) & (3) & (4)\\ 
\hline \\[-1.8ex] 
 YEAR2020 & 0.562 & 0.999 & 0.577 & 0.729 \\ 
  & (0.340, 0.930) & (0.607, 1.645) & (0.349, 0.954) & (0.436, 1.221) \\ 
  & p = 0.025$^{*}$ & p = 0.999 & p = 0.033$^{*}$ & p = 0.230 \\ 
  & & & & \\ 
 GENDERMale:YEAR2020 & 1.362 & 1.717 & 1.265 & 1.563 \\ 
  & (1.001, 1.852) & (1.264, 2.332) & (0.929, 1.723) & (1.127, 2.167) \\ 
  & p = 0.050$^{*}$ & p = 0.001$^{**}$ & p = 0.136 & p = 0.008$^{**}$ \\ 
  & & & & \\ 
 AGECAT30-49:YEAR2020 & 1.203 & 0.813 & 1.468 & 2.119 \\ 
  & (0.782, 1.851) & (0.530, 1.248) & (0.954, 2.257) & (1.369, 3.279) \\ 
  & p = 0.400 & p = 0.345 & p = 0.081 & p = 0.001$^{**}$ \\ 
  & & & & \\ 
 AGECAT50-64:YEAR2020 & 0.785 & 0.408 & 1.829 & 1.650 \\ 
  & (0.493, 1.248) & (0.257, 0.649) & (1.146, 2.920) & (1.015, 2.683) \\ 
  & p = 0.307 & p = 0.0002$^{**}$ & p = 0.012$^{*}$ & p = 0.044$^{*}$ \\ 
  & & & & \\ 
 AGECAT65+:YEAR2020 & 0.695 & 0.406 & 1.890 & 2.141 \\ 
  & (0.417, 1.157) & (0.245, 0.674) & (1.135, 3.146) & (1.261, 3.636) \\ 
  & p = 0.162 & p = 0.0005$^{**}$ & p = 0.015$^{*}$ & p = 0.005$^{**}$ \\ 
  & & & & \\ 
 RACEBlack or African American:YEAR2020 & 0.835 & 1.101 & 0.783 & 0.781 \\ 
  & (0.535, 1.303) & (0.707, 1.715) & (0.499, 1.228) & (0.489, 1.248) \\ 
  & p = 0.427 & p = 0.671 & p = 0.287 & p = 0.302 \\ 
  & & & & \\ 
 RACEAsian or Asian-American:YEAR2020 & 0.729 & 1.059 & 0.672 & 0.871 \\ 
  & (0.366, 1.451) & (0.530, 2.116) & (0.353, 1.279) & (0.443, 1.712) \\ 
  & p = 0.368 & p = 0.871 & p = 0.227 & p = 0.689 \\ 
  & & & & \\ 
 RACEOther:YEAR2020 & 1.524 & 1.140 & 1.070 & 1.074 \\ 
  & (0.885, 2.624) & (0.663, 1.958) & (0.618, 1.851) & (0.598, 1.929) \\ 
  & p = 0.129 & p = 0.636 & p = 0.811 & p = 0.811 \\ 
  & & & & \\ 
 HISPANICYes:YEAR2020 & 0.837 & 0.820 & 0.524 & 0.914 \\ 
  & (0.513, 1.364) & (0.503, 1.335) & (0.318, 0.863) & (0.543, 1.539) \\ 
  & p = 0.475 & p = 0.425 & p = 0.012$^{*}$ & p = 0.735 \\ 
  & & & & \\ 
 EDUCATIONBachelor's or more:YEAR2020 & 1.586 & 1.259 & 1.824 & 1.706 \\ 
  & (1.100, 2.286) & (0.875, 1.811) & (1.263, 2.635) & (1.158, 2.515) \\ 
  & p = 0.014$^{*}$ & p = 0.215 & p = 0.002$^{**}$ & p = 0.007$^{**}$ \\ 
  & & & & \\ 
 EDUCATIONSome college:YEAR2020 & 1.412 & 1.005 & 1.329 & 1.475 \\ 
  & (0.961, 2.076) & (0.685, 1.475) & (0.903, 1.956) & (0.982, 2.214) \\ 
  & p = 0.079 & p = 0.979 & p = 0.149 & p = 0.062 \\ 
  & & & & \\ 
 POLPARTYRep/Lean Rep:YEAR2020 & 0.940 & 1.129 & 1.239 & 1.176 \\ 
  & (0.680, 1.297) & (0.818, 1.558) & (0.894, 1.716) & (0.832, 1.662) \\ 
  & p = 0.706 & p = 0.462 & p = 0.199 & p = 0.359 \\ 
  & & & & \\ 
\hline \\[-1.8ex] 
Observations & 3,150 & 3,150 & 3,127 & 3,127 \\ 
Log Likelihood & $-$2,130.834 & $-$2,130.678 & $-$2,088.554 & $-$1,854.825 \\ 
Akaike Inf. Crit. & 4,309.667 & 4,309.355 & 4,225.108 & 3,757.649 \\ 
\hline 
\hline \\[-1.8ex] 
\textit{Note:}  & \multicolumn{4}{r}{$^{*}$p$<$0.05; $^{**}$p$<$0.01} \\ 
\end{tabular} }
\end{table*}

\begin{table*}[!htbp] \centering 
  \caption{Interaction between 2020 and 2021, with 2020 as reference. Fixed effects are omitted for brevity. See Table~\ref{tbl:QuestionA_years}-\ref{tbl:QuestionE_years} for reference.} 
  \label{tbl:model_2020_2021} 
  \resizebox{\linewidth}{11cm}{
\begin{tabular}{@{\extracolsep{5pt}}lcccc} 
\\[-1.8ex]\hline 
\hline \\[-1.8ex]

 & Question 1 & Question 2 & Question 3 & Question 4 \\ 
 & Government & Law Enforcement & Medical Research & Corporate \\ 
 & Terrorism & Genetic & Fitness & Mental Health \\ 
\\[-1.8ex] & (1) & (2) & (3) & (4)\\ 
\hline \\[-1.8ex] 
 YEAR2021 & 1.153 & 0.750 & 1.133 & 1.175 \\ 
  & (0.692, 1.920) & (0.454, 1.239) & (0.682, 1.881) & (0.709, 1.948) \\ 
  & p = 0.585 & p = 0.262 & p = 0.630 & p = 0.532 \\ 
  & & & & \\ 
 GENDERMale:YEAR2021 & 0.974 & 0.828 & 1.121 & 1.032 \\ 
  & (0.703, 1.350) & (0.600, 1.144) & (0.810, 1.552) & (0.744, 1.432) \\ 
  & p = 0.874 & p = 0.253 & p = 0.491 & p = 0.850 \\ 
  & & & & \\ 
 AGECAT30-49:YEAR2021 & 0.979 & 1.160 & 0.812 & 1.006 \\ 
  & (0.639, 1.499) & (0.762, 1.767) & (0.531, 1.241) & (0.661, 1.530) \\ 
  & p = 0.923 & p = 0.488 & p = 0.336 & p = 0.980 \\ 
  & & & & \\ 
 AGECAT50-64:YEAR2021 & 1.126 & 1.592 & 1.401 & 0.980 \\ 
  & (0.686, 1.848) & (0.975, 2.598) & (0.856, 2.293) & (0.593, 1.619) \\ 
  & p = 0.638 & p = 0.064 & p = 0.181 & p = 0.936 \\ 
  & & & & \\ 
 AGECAT65+:YEAR2021 & 0.975 & 1.526 & 1.524 & 0.883 \\ 
  & (0.568, 1.671) & (0.900, 2.588) & (0.894, 2.597) & (0.518, 1.506) \\ 
  & p = 0.926 & p = 0.117 & p = 0.122 & p = 0.649 \\ 
  & & & & \\ 
 RACEBlack or African American:YEAR2021 & 1.397 & 1.238 & 0.944 & 0.882 \\ 
  & (0.898, 2.176) & (0.799, 1.916) & (0.607, 1.466) & (0.567, 1.373) \\ 
  & p = 0.139 & p = 0.339 & p = 0.797 & p = 0.580 \\ 
  & & & & \\ 
 RACEAsian or Asian-American:YEAR2021 & 0.898 & 1.072 & 1.490 & 0.921 \\ 
  & (0.526, 1.535) & (0.627, 1.831) & (0.868, 2.557) & (0.538, 1.575) \\ 
  & p = 0.695 & p = 0.800 & p = 0.149 & p = 0.763 \\ 
  & & & & \\ 
 RACEOther:YEAR2021 & 0.808 & 1.144 & 0.907 & 1.047 \\ 
  & (0.439, 1.488) & (0.627, 2.087) & (0.495, 1.663) & (0.569, 1.926) \\ 
  & p = 0.495 & p = 0.662 & p = 0.753 & p = 0.884 \\ 
  & & & & \\ 
 HISPANICYes:YEAR2021 & 1.844 & 1.460 & 1.443 & 1.425 \\ 
  & (1.021, 3.331) & (0.813, 2.624) & (0.799, 2.606) & (0.788, 2.579) \\ 
  & p = 0.043$^{*}$ & p = 0.206 & p = 0.225 & p = 0.242 \\ 
  & & & & \\ 
 EDUCATIONBachelor's or more:YEAR2021 & 0.888 & 1.071 & 0.852 & 0.777 \\ 
  & (0.597, 1.322) & (0.723, 1.585) & (0.572, 1.268) & (0.522, 1.158) \\ 
  & p = 0.561 & p = 0.734 & p = 0.430 & p = 0.217 \\ 
  & & & & \\ 
 EDUCATIONSome college:YEAR2021 & 0.825 & 1.207 & 0.958 & 0.773 \\ 
  & (0.547, 1.245) & (0.804, 1.812) & (0.635, 1.444) & (0.512, 1.166) \\ 
  & p = 0.360 & p = 0.363 & p = 0.837 & p = 0.220 \\ 
  & & & & \\ 
 POLPARTYRep/Lean Rep:YEAR2021 & 0.522 & 0.642 & 0.497 & 0.886 \\ 
  & (0.371, 0.735) & (0.458, 0.902) & (0.352, 0.700) & (0.628, 1.249) \\ 
  & p = 0.0003$^{**}$ & p = 0.011$^{*}$ & p = 0.0001$^{**}$ & p = 0.489 \\ 
  & & & & \\ 
\hline \\[-1.8ex] 
Observations & 2,675 & 2,675 & 2,675 & 2,675 \\ 
Log Likelihood & $-$1,796.424 & $-$1,827.617 & $-$1,800.311 & $-$1,795.949 \\ 
Akaike Inf. Crit. & 3,640.848 & 3,703.234 & 3,648.622 & 3,639.897 \\ 
\hline 
\hline \\[-1.8ex] 
\textit{Note:}  & \multicolumn{4}{r}{$^{*}$p$<$0.05; $^{**}$p$<$0.01} \\ 
\end{tabular} }
\end{table*}

\begin{table*}[!htbp] \centering 
  \caption{Interaction between 2019 and 2021, with 2019 as reference. Fixed effects are omitted for brevity. See Table~\ref{tbl:QuestionA_years}-\ref{tbl:QuestionE_years} for reference.} 
  \label{tbl:model_2019_2021} 
  \resizebox{\linewidth}{11cm}{
\begin{tabular}{@{\extracolsep{5pt}}lcccc} 
\\[-1.8ex]\hline 
\hline \\[-1.8ex]

 & Question 1 & Question 2 & Question 3 & Question 4 \\ 
 & Government & Law Enforcement & Medical Research & Corporate \\ 
 & Terrorism & Genetic & Fitness & Mental Health \\ 
\\[-1.8ex] & (1) & (2) & (3) & (4)\\ 
\hline \\[-1.8ex] 
 YEAR2021 & 0.649 & 0.750 & 0.653 & 0.857 \\ 
  & (0.423, 0.994) & (0.490, 1.147) & (0.425, 1.005) & (0.551, 1.332) \\ 
  & p = 0.047$^{*}$ & p = 0.184 & p = 0.053 & p = 0.494 \\ 
  & & & & \\ 
 GENDERMale:YEAR2021 & 1.326 & 1.422 & 1.418 & 1.614 \\ 
  & (1.002, 1.755) & (1.076, 1.879) & (1.069, 1.881) & (1.195, 2.180) \\ 
  & p = 0.049$^{*}$ & p = 0.014$^{*}$ & p = 0.016$^{*}$ & p = 0.002$^{**}$ \\ 
  & & & & \\ 
 AGECAT30-49:YEAR2021 & 1.178 & 0.944 & 1.191 & 2.131 \\ 
  & (0.801, 1.733) & (0.642, 1.387) & (0.808, 1.756) & (1.439, 3.156) \\ 
  & p = 0.406 & p = 0.768 & p = 0.376 & p = 0.0002$^{**}$ \\ 
  & & & & \\ 
 AGECAT50-64:YEAR2021 & 0.884 & 0.650 & 2.562 & 1.616 \\ 
  & (0.574, 1.361) & (0.423, 0.999) & (1.657, 3.962) & (1.024, 2.551) \\ 
  & p = 0.575 & p = 0.050$^{*}$ & p = 0.00003$^{**}$ & p = 0.040$^{*}$ \\ 
  & & & & \\ 
 AGECAT65+:YEAR2021 & 0.677 & 0.620 & 2.880 & 1.891 \\ 
  & (0.428, 1.070) & (0.394, 0.975) & (1.819, 4.560) & (1.171, 3.054) \\ 
  & p = 0.096 & p = 0.039$^{*}$ & p = 0.00001$^{**}$ & p = 0.010$^{**}$ \\ 
  & & & & \\ 
 RACEBlack or African American:YEAR2021 & 1.166 & 1.363 & 0.739 & 0.689 \\ 
  & (0.765, 1.778) & (0.893, 2.078) & (0.479, 1.139) & (0.439, 1.080) \\ 
  & p = 0.475 & p = 0.151 & p = 0.170 & p = 0.105 \\ 
  & & & & \\ 
 RACEAsian or Asian-American:YEAR2021 & 0.655 & 1.135 & 1.001 & 0.802 \\ 
  & (0.335, 1.279) & (0.579, 2.225) & (0.536, 1.872) & (0.417, 1.542) \\ 
  & p = 0.216 & p = 0.712 & p = 0.997 & p = 0.508 \\ 
  & & & & \\ 
 RACEOther:YEAR2021 & 1.232 & 1.304 & 0.970 & 1.124 \\ 
  & (0.689, 2.203) & (0.736, 2.309) & (0.543, 1.734) & (0.609, 2.075) \\ 
  & p = 0.482 & p = 0.364 & p = 0.919 & p = 0.708 \\ 
  & & & & \\ 
 HISPANICYes:YEAR2021 & 1.543 & 1.197 & 0.756 & 1.303 \\ 
  & (0.914, 2.606) & (0.712, 2.012) & (0.444, 1.286) & (0.750, 2.262) \\ 
  & p = 0.105 & p = 0.498 & p = 0.303 & p = 0.348 \\ 
  & & & & \\ 
 EDUCATIONBachelor's or more:YEAR2021 & 1.409 & 1.348 & 1.554 & 1.327 \\ 
  & (0.999, 1.987) & (0.958, 1.897) & (1.098, 2.200) & (0.920, 1.913) \\ 
  & p = 0.051 & p = 0.087 & p = 0.013$^{*}$ & p = 0.131 \\ 
  & & & & \\ 
 EDUCATIONSome college:YEAR2021 & 1.165 & 1.214 & 1.273 & 1.139 \\ 
  & (0.817, 1.661) & (0.854, 1.726) & (0.890, 1.820) & (0.782, 1.659) \\ 
  & p = 0.399 & p = 0.282 & p = 0.187 & p = 0.497 \\ 
  & & & & \\ 
 POLPARTYRep/Lean Rep:YEAR2021 & 0.490 & 0.725 & 0.615 & 1.041 \\ 
  & (0.365, 0.659) & (0.541, 0.972) & (0.456, 0.830) & (0.758, 1.431) \\ 
  & p = 0.00001$^{**}$ & p = 0.032$^{*}$ & p = 0.002$^{**}$ & p = 0.803 \\ 
  & & & & \\ 
\hline \\[-1.8ex] 
Observations & 3,549 & 3,549 & 3,526 & 3,526 \\ 
Log Likelihood & $-$2,388.064 & $-$2,402.085 & $-$2,340.825 & $-$2,119.949 \\ 
Akaike Inf. Crit. & 4,824.129 & 4,852.170 & 4,729.651 & 4,287.898 \\ 
\hline 
\hline \\[-1.8ex] 
\textit{Note:}  & \multicolumn{4}{r}{$^{*}$p$<$0.05; $^{**}$p$<$0.01} \\ 
\end{tabular} }
\end{table*}

\end{document}